\documentclass{jfm}
\usepackage[version=3]{mhchem}
\usepackage{subfigure}
\usepackage{color}
 \usepackage{dblfloatfix}
 \usepackage{graphicx}

\usepackage{citesort}
\usepackage{amssymb}
\usepackage{amsmath}
\usepackage{setspace}
\usepackage{siunitx}
\usepackage{url,hyperref}
\DeclareMathAlphabet{\mathpzc}{OT1}{pzc}{m}{it}

\usepackage{amsfonts}
\usepackage{physics}
\hypersetup{
    colorlinks = true,
    urlcolor   = blue,
    citecolor  = black,
}

\newcommand{\RomanNumeralCaps}[1]
\nolinenumbers
\newcommand{\vv}{\mbox{\boldmath $v$}}
\newcommand{\vn}{\mbox{\boldmath $n$}}
\newcommand{\vtau}{\mbox{\boldmath $\tau$}}
\newcommand{\vz}{\mbox{\boldmath $e_z$}}

\newcommand{\E}{\mathcal E}
\newcommand{\Pe}{\mbox{\rm Pe}}
\newcommand{\Rere}{\mbox{\rm Re}}
\newcommand{\Prpr}{\mbox{\rm Pr}}
\renewcommand{\L}{\mathcal L}

\newcommand{\Gr}{\mbox{\rm Gr}}
\newcommand{\ddd}{\mathrm d}

\title{Small Leidenfrost droplet dynamics}

\author{Benjamin Sobac\aff{1}
  \corresp{\email{benjamin.sobac@cnrs.fr}},
  Alexey Rednikov\aff{2}
 \and Pierre Colinet\aff{2}}

\affiliation{\aff{1} CNRS, Universite de Pau et des Pays de l'Adour, E2S UPPA, LFCR, 64600 Anglet, France
\aff{2}TIPs Lab, Universite libre de Bruxelles, 1050 Brussels, Belgium
}

\begin{document}
\maketitle

\begin{abstract}
An isolated Leidenfrost droplet levitating over its own vapor above a superheated flat substrate is considered theoretically, the superheating for water being up to several hundred degrees above the boiling temperature. The focus is on the limit of small, practically spherical droplets of several tens of micrometers or less. This may occur when the liquid is sprayed over a hot substrate, or just be a late life stage of an initially large Leidenfrost droplet. A rigorous numerically-assisted analysis is carried out within verifiable assumptions such as quasi-stationarities and small Reynolds/P\'{e}clet numbers. It is considered that the droplet is surrounded by its pure vapor. Simple fitting formulas of our numerical data for the forces and evaporation rates are preliminarily obtained, all respecting the asymptotic behaviors (also investigated) in the limits of small and large levitation heights. They are subsequently used within a system of ODEs to study the droplet dynamics and take-off (drastic height increase as the droplet vaporizes). A previously known quasi-stationary inverse-square root law for the droplet height as a function of its radius (at the root of the take-off) is recovered, although we point out different prefactors in the two limits. Deviations of a dynamic nature therefrom are uncovered as the droplet radius further decreases due to evaporation, improving the agreement with experiment. Furthermore, we reveal that, if initially large enough, the droplets vanish at a universal finite height (just dependent on the superheat and fluid properties). Scalings in various distinguished cases are obtained in the way.  
\end{abstract}

\section{Introduction}
\label{sec:intro}

When a volatile liquid droplet is placed on a hot solid surface, superheated well above the boiling temperature, it neither touches the substrate nor boils, but rather floats on a thin film of its own vapor. This fascinating phenomenon, known as the \textit{Leidenfrost effect}, does not cease to attract attention since its first descriptions about 300 years ago \citep{Boerhaave1732,1756Leidenfrost}. This is due to not only the myriad of intriguing and unexpected behaviors a droplet can exhibit in this state, 
but also its relevance across a wide range of industrial and technological processes, spanning from the traditional heat transfer applications to the emerging field of multiphase milli-/micro-fluidics. See, for example, the review articles~\citep{2013_Quere,2021_Ajaev,2022_Stewart}, dedicated book chapters in ~\cite{2015_Brutin,2022_Marengo}, and the many references therein.

The vapor film, a key feature of the Leidenfrost state, ensures the droplet levitation while acting as a thermal insulator, resulting in relatively low evaporation rates and hence long lifetimes of the droplet. In this state, the droplet's weight is balanced by the pressure within such a vapor cushion squeezed by the slowly and steadily evaporating drop. With no contact with the substrate, the observable shapes of the droplets are governed by a balance between capillarity and gravity similarly to a perfectly non-wetting (superhydrophobic) situation. Denoting the capillary length by $\ell_c$, droplets with radii $R$ smaller than $\ell_c$ remain quasi-spherical while puddles larger than $\ell_c$ are flattened by gravity, whose height is limited by $\approx2\ell_c$~\citep{2003_Biance}. The profile of the underlying vapor film is non-trivial. For a large droplet with $R\gtrsim \ell_c$, the vapor film exhibits a pocket-like structure composed by an internal vapor `pocket' surrounded by a thin neck. As the drop gets smaller, the vapor film slimes down, the drop getting closer to the substrate. When the droplet radius is small enough as compared to $\ell_c$, the vapor pocket disappears completely, and the droplet becomes quasi-spherical with a small circular area slightly flattened at the bottom. Accurate interferometric measurements of the vapor film thickness profile~\citep{2012_Burton} turn out to be in a good agreement with a refined theoretical modeling ~\citep{2014_Sobac,2021_Sobac_Erratum} coupling lubricated vapor flow, capillarity and hydrostatic pressure effects, itself recently confirmed numerically by \cite{2022_Chakraborty}. Note that the main scaling laws featuring the shapes of a Leidenfrost droplet and its evaporation dynamics can be found in \cite{2003_Biance,2012_Pomeau,2014_Sobac,2021_Sobac_Erratum}.

In practice, this absence of contact between the Leidenfrost droplet and the substrate leads to very rich dynamics. For large puddle-like drops, the vapor pocket grows until it eventually pops up as a central `chimney' due to a Rayleigh-Taylor mechanism~\citep{2003_Biance, 2009_Snoeijer}. Instability of large droplets can also occur (either spontaneously or forced) in the form of `star-faceted' shapes when azimuthal surface oscillations develop along the periphery of the droplets~\citep{2011_Brunet, 2017_Ma, 2018_Ma, 2019_Bergen, 2021_Bouillant_PNAS}. Self-induced spontaneous oscillations can also occurs in the vertical plane yielding to the recently reported bobing, bouncing or trampolining dynamics when Leidenfrost droplets reach moderate and small size with $R\leqslant\ell_c$ ~\citep{2020_Liu, 2021_Graeber}. Other spectacular behaviors related to their high mobility have been observed. These include Leidenfrost wheels, when a droplet initially at rest spontaneously rolls and moves over a flat surface like a wheel due to symmetry breaking in the internal flow of the liquid~\citep{2018_Bouillant}, and self-propelling of Leidenfrost droplets when interacting with substrate breaking the axi-symmetry, either due to surface topography such as ratchets or herringbones~\citep{2006_Linke, 2011_Dupeux, 2012_Marin, 2013_Baier, 2016_Soto, 2019_Dodd}, or temperature gradients~\citep{2017_Sobac, 2020_Dodd, 2021_Bouillant_SM}. Thus, droplets move in a direction dictated by the patterns due to symmetry breaking of the vapor layer.  Strategies have emerged to control the motion and manipulate these droplets. In addition to geometric and thermal heterogeneities, chemical patterns of the surface can also be exploited to tailor the vapor film, enabling the stretching, sloshing, spinning, propelling, or trapping of a Leidenfrost droplet~\citep{2023_Li}.  

As compared to large and moderate-size, the dynamics of small, near-spherical Leidenfrost drops  ($R\ll\ell_c$) has not received so much attention.  In their seminal work, \cite{2012_Celestini} first explored the final fate of Leidenfrost drops as they became very small, just moments before disappearing. By spraying tiny droplets of water or ethanol in the size range of about $1-30\,\mu\mathrm{m}$  onto a superheated substrate, they discovered that when small enough (i.e, with $R$ below a characteristic radius corresponding to the breakup of the lubrication approximation), Leidenfrost droplets took off from the heated substrate with an elevation $h\propto R^{-1/2}$, as predicted by \cite{2012_Celestini} and also by \cite{2012_Pomeau}. Remarkably, in this regime, droplets become too light to stand over the upward force generated by the pressure due to evaporation, and they reach higher and higher elevations while vaporizing. This behavior drastically contrasts with what is observed for larger Leidenfrost droplets. More recently, \cite{2019_Lyu} observed that a second final fate, other than lift-off, is possible for Leidenfrost droplets. Namely, if the liquid droplet is not pure or contaminant-free enough, small Leidenfrost droplets are unable to take off, but instead disappear by exploding with an audible crack.

Here, we propose to theoretically revisit the dynamics of small spherical Leidenfrost droplets with the aim of comprehensively and thoroughly analyzing the mechanisms involved in their final fate. Thanks to a model including a realistic description of the coupling between hydrodynamics, heat transfer and evaporation, this work seems to be the first to provide exact estimates of the drop elevation as a function of the physical parameters and without any fitting parameter. After numerically computing the entirety of fluxes, evaporation rates and forces, a master curve for droplet elevation as a function of its size is derived by simply balancing the drop weight with the upward evaporation-induced hydrodynamic force. While the scaling law agrees with \cite{2012_Celestini,2012_Pomeau}, there appear subtleties concerning the prefactor. Moreover, the analysis reveals that such a classical quasi-steady description is not fully sufficient to describe the take-off phenomenon. Even at these small scales, further dynamical effects must be taken into account to achieve a good agreement with the original experimental data of \cite{2012_Celestini}.

\section{Statement of the problem, premises and outlook}
\label{formul}

\begin{figure}
  \centerline{\includegraphics[width=0.4\columnwidth]{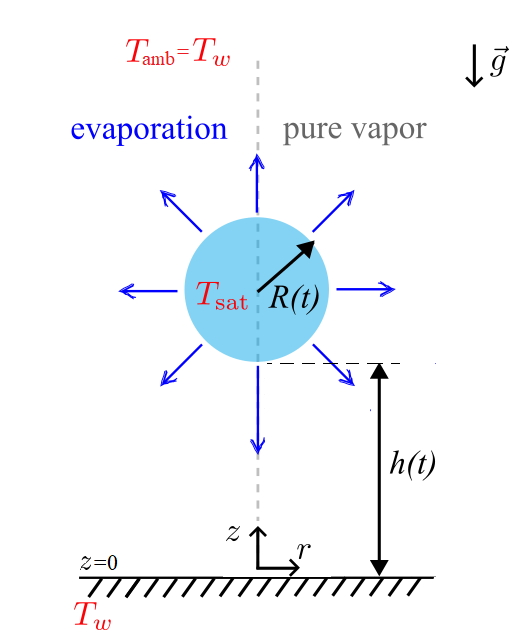}}
  \caption{Sketch of the problem. 
  }
\label{fig1}
\end{figure}

Consider a small evaporating spherical droplet of radius $R$ in a Leidenfrost state levitating at a height $h$ above a superheated substrate at a `wall' temperature $T_w$, as sketched in Fig.~\ref{fig1}. The substrate is flat and horizontal. We shall be interested in the take-off of Leidenfrost droplets \citep{2012_Celestini,2012_Pomeau}, which occurs in the realm of small droplets ($R$ of the order of tens of $\mu$m) with a negligible deviation from the spherical shape.  The (immediate) surroundings of the droplet are assumed to be saturated with vapour (totally displacing the air) and heated through to the substrate temperature. Thus, $T_w$ is here also an effective overall ambient temperature (i.e.~$T_\text{amb}=T_w$). The droplet is assumed isothermal at saturation (boiling) temperature $T_\text{sat}$. The superheat is given by $\Delta T=T_w-T_{\rm{sat}}>0$. The typical values considered here are $T_\text{sat}=100^\circ\text{C}$ (water at atmospheric pressure), $T_w=400^\circ\text{C}$, and $\Delta T=300^\circ\text{C}$. Such superheat occurs e.g.\ in the experiments by~\cite{2012_Celestini}.

Mathematically, the goal of the present consideration is obtaining an interrelation between $h$, $R$ (in particular, as functions of time $t$ due to the droplet evaporation) and the parameters of the problem, such as $\Delta T$, $g$ (gravity acceleration, $9.81\,\mathrm{m/s^2}$) and the liquid and vapour properties. The latter include $\rho_l$ (liquid density, defined at $T_\text{sat}$) as well as the following vapour properties: $\rho_v$ (density), $\mu_v$ (dynamic viscosity), $\nu_v=\mu_v/\rho_v$ (kinematic viscosity), $\lambda_v$ (thermal conductivity), $c_{p,v}$ (heat capacity at constant pressure), $\alpha_v=\lambda_v/(\rho_v c_{p,v})$ (thermal diffusivity), $\L$ (latent heat of evaporation). These may vary considerably in the temperature range between $T_\text{sat}$ and $T_w$. 
However, for the sake of simplicity, we shall here assume them constant and defined at the mid-temperature $\frac{1}{2}(T_w+T_\text{sat})$, except for $\L$ defined at $T_\text{sat}$, similarly to the approach used elsewhere~\citep{2014_Sobac,2015_Sobac}. The relevant property values are provided in~Appendix~\ref{app:Properties}.

Other key assumptions include negligible advective/convective effects (small P\'eclet and Reynolds numbers), so that the temperature field in the vapour is governed by heat conduction, while the evaporative flow from the droplet can be considered by means of the Stokes approximation. 
Quasi-steadiness of the temperature and velocity fields, in spite of $R$ and $h$ changing in time due to evaporation, is another key assumption of the analysis.  
In other words, these fields and the evaporation fluxes and forces they determine are merely functions of the instantaneous values of $R$ and $h$ and do not depend on the history. It is under this premise that a preliminary calculation of these quantities is carried out in \S\ref{basics}. The validity of this and other assumptions is verified \textit{a~posteriori} in their due course.

The quasi-steadiness assumption is also applied at first when it comes to the force balance on the droplet in \S\ref{quasi-sta_lev}, permitting to predict the levitation height $h$ and make a first comparison with experiment. Yet, certain limitations are thereby disclosed, inspiring consideration of a more general droplet dynamics in \S\ref{basics2}--\S\ref{dyn_gen}. 
However, even in such a situation, the quasi-steadiness of the quantities like calculated in \S\ref{basics} is still assumed to hold.

An important geometric parameter of the configuration (figure~\ref{fig1}), meriting a special notation, is the ratio of the droplet's height and radius: 
\begin{equation}
\delta=\frac{h}{R}\,.
\label{ratio}
\end{equation}
In the present study, we shall be interested in a full range of this relative-height' parameter, from very small to very large. The large values are expected for small droplets (small $R$) upon a take-off~\citep{2012_Celestini,2012_Pomeau,2021_Sobac_Erratum,2022_Chakraborty}. In contrast, small $\delta$ are attained for larger droplets. In this way, we arrive at a transition from spherical Leidenfrost droplets to Leidenfrost droplets for which deformation (at first at the bottom slightly flattened by gravity) becomes essential. Such a transition is touched upon in \S\ref{global}.

\section{Basic calculations: fields, fluxes and forces}
\label{basics}

\begin{table*}
      
\begin{tabular}{ccccccccc}
  \hline
  $[r,z,s]$ & $[S]$ & $[T-T_\text{sat}]$ & $[j]$  &  $[J]$ & $[\vv]$ & $[p]$  & $[F_\text{ev}]$ \\
  \hline
 $R$ & $R^2$ & $\Delta T$ & $\frac{\lambda_v \Delta T}{\mathcal{L} R}$ & $\quad [j] R^2=\frac{\lambda_v \Delta T R}{\mathcal{L}}$  & $\quad\frac{[j]}{\rho_v}=\frac{\lambda_v \Delta T}{\rho_v \mathcal{L} R}$ & $\quad\mu_v \frac{[\vv]}{R}$ & $\quad\mu_v R [\vv]=\frac{\mu_v\lambda_v \Delta T}{\rho_v \mathcal{L}}$ \\ 
  \hline
  \end{tabular}
\caption{Scales used to render the various quantities dimensionless in \S\ref{basics}. The square brackets denote the scale of the quantity inside.}
\label{tab:scales}
   \end{table*}

Dimensionless variables are introduced using the scales given in table~\ref{tab:scales} (definitions to be given in their due course). For simplicity and expecting no confusion, no notation distinction is made between the original, dimensional variables and their dimensionless versions in the present section (the distinction being clear from the context). We just note that 
a dimensionless temperature is introduced as 
\begin{equation}
\hat{T}=\frac{T-T_\text{sat}}{\Delta T}
\label{Tadimdef}
\end{equation}
where recall that $\Delta T=T_w-T_\text{sat}$. Hereafter, in the same spirit, the hats are omitted for the sake of brevity.

\subsection{Temperature field}
\label{Tfield}

As stipulated in \S~\ref{basics}, 
the heat transport in the gas phase is conductive and quasi-steady. Thus, the thermal problem is decoupled from the evaporative velocity field, and the dimensionless temperature field $T$ is governed by the Laplace equation
\begin{equation}
\nabla^2{T}=0.
\label{eq:lapl}
\end{equation}
\noindent It is subject to the boundary conditions:
\begin{eqnarray}
T&=&1 \quad \text{on the hot substrate,}\label{eq:BCT1}\\
T&\to&1 \quad \text{far away from the drop,}\label{eq:BCT2}\\
T&=&0 \quad \text{on the droplet surface.} \label{eq:BCT3}
\end{eqnarray}
Although an exact solution in bipolar coordinates can be found e.g.\ using the methods by~\cite{Lebedev}, it is rather cumbersome so that we eventually opt for a numerical solution using COMSOL Multiphysics.

\begin{figure}
  \centerline{\includegraphics[width=\columnwidth]{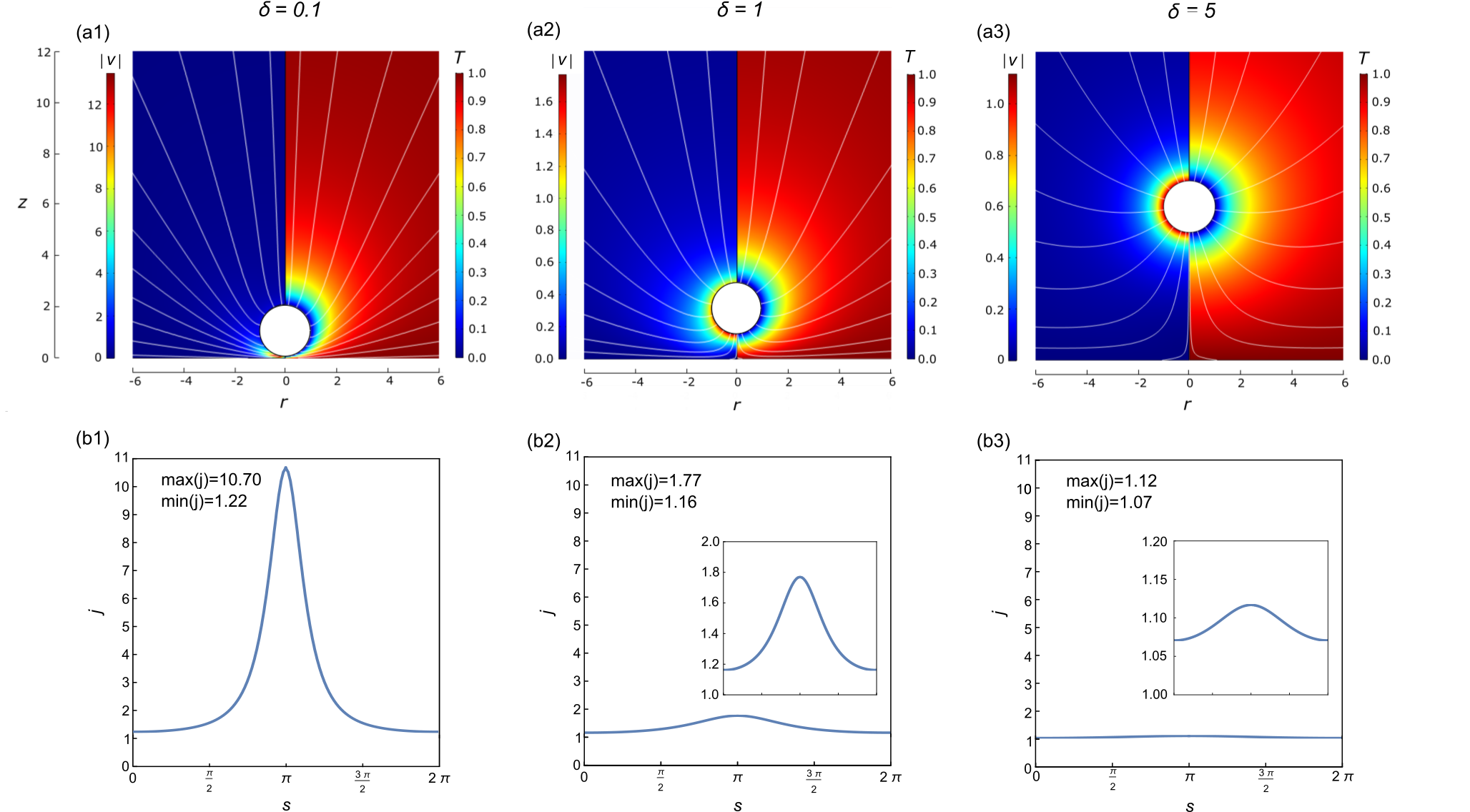}}
  \caption{
  (a) Dimensionless temperature (right) and velocity (left) fields for various values of $\delta$. Streamlines are also shown by white lines. 
  (b) Corresponding profiles of the dimensionless evaporation flux $j$ along the droplet surface as a function of the dimensionless arclength $s$ ($s=0$ at the droplet apex, $s=\pi$ at its lowest point; the plotting is formally continued up to $s=2\pi$ back to the apex for aesthetic purposes). Inserts are simply zooms of the main plot.}
\label{fig2}
\end{figure}

The results of the simulations are shown in Fig.~\ref{fig2}(a) for three different values of the separating distance $\delta$: a large droplet very close to the substrate with $\delta= 0.1$; a droplet at a distance from the substrate comparable to its radius with $\delta= 1$, and a small droplet beginning to be far away from the substrate with $\delta=5$. One immediately observes that at small $\delta$, the temperature difference is squeezed into a thin film between the droplet and the substrate. At large $\delta$, the temperature field approaches a spherically symmetric one, as expected. Other results displayed in figure~\ref{fig2} will be discussed later.

\subsection{Evaporation rate}
\label{Jevap}

At the droplet surface, the evaporation mass flux $j$ $[\mathrm{kg/m^2 s}]$ is at the expense of the heat coming by from the superheated surroundings through the vapor phase: $j=\frac{\lambda_v}{\mathcal{L}} \vn\cdot\grad T$, where $\vn$ is the (external) unit normal vector. In dimensionless terms (cf.~table~\ref{tab:scales}), this reads 
\begin{equation}
\label{eq:evaporation_flux}
j =  \vn \cdot \grad{T} \,.
\end{equation}
Using the temperature field computed in \S~\ref{Tfield}, the profiles of the evaporation flux $j$ are calculated from (\ref{eq:evaporation_flux}) and shown in Fig.~\ref{fig2}(b). One can appreciate that due to the presence of the hot substrate, $j$ is maximum at the base and decreases towards the apex, where the minimum is attained. The closer the droplet is to the substrate (the smaller $\delta$ is), the more the profile of $j$ is non-uniform and the values of $j$ are large.  
At small $\delta$, one obviously obtains max($j$) $\propto 1/\delta$ (heat conduction across a thin vapour layer). When the relative droplet height $\delta$ increases, the non-uniformity of $j$ weakens and the average of $j$ decreases. Eventually, $j$ tends to a uniform value of 1 for $\delta\gg 1$ (as for a droplet in an unbounded medium). 

\begin{figure}
  \centerline{\includegraphics[width=0.65\columnwidth]{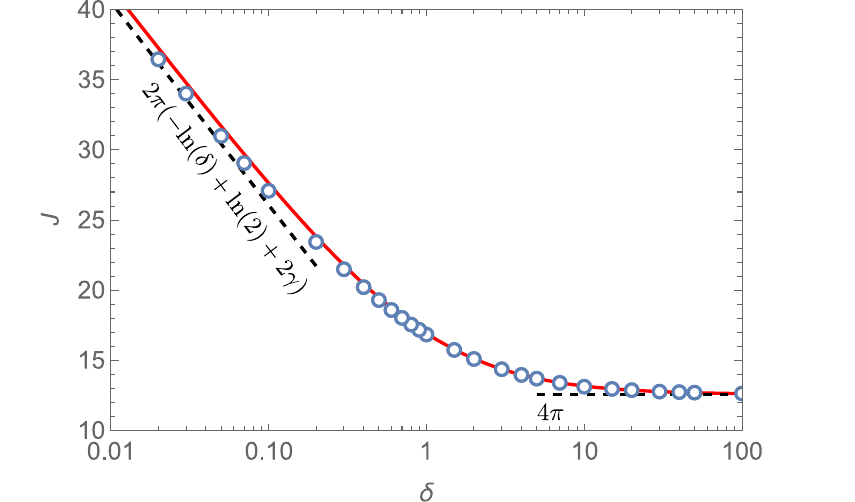}}
  \caption{Evaporation rate $J$ (dimensionless) as a function of the relative droplet height $\delta$. Numerical results (blue open circles) are fairly well approximated by equation~(\ref{eq:FitJbis}) (red line). Black dashed lines correspond to the asymptotic behaviours (Appendix~\ref{app:Asymptotic}).}
\label{fig3}
\end{figure}

The (global) evaporation rate $J$ can be directly deduced by integrating the evaporation flux all over the droplet surface:
\begin{equation}
 J=\iint j\, \mathtt{d}S \,.
\end{equation}
Figure~\ref{fig3} reports the computed values of $J$ as a function of $\delta$. As expected from the knowledge of the $j$ behaviour,  $J$ diverges as $\delta\to 0$ and decreases to saturate at $4\pi$ as $\delta\to +\infty$. Such asymptotic behaviours are investigated in detail in Appendices~\ref{Jasy0} and~\ref{Jasyinf} and also represented in figure~\ref{fig3}. A good simple fit of the numerical data for $J$, respecting the leading-order asymptotic behaviours, is given by 
\begin{equation}
J(\delta)=4\pi \left[ 1+\frac{1}{2} \ln\left(1+\frac{1}{\delta}\right) \right]\,,
\label{eq:FitJbis}
\end{equation}
where a maximum deviation from the data does not exceed 2.7\%.  
However, a more precise fit is also provided for reference in Appendix~\ref{app:Jfit}. 
   
\subsection{Velocity field}
\label{sec:velocity}
In accordance with the approach followed in the present paper (\S~\ref{formul}), the evaporative flow field, generated by droplet evaporation, is considered in the Stokes and quasi-steady approximation. Thus, we proceed from the continuity and Stokes equations 
\begin{eqnarray}
\div{\vv}&=&0,\label{eq:stokes1}\\
\nabla^2{{\vv}}-\grad{p}&=&0. 
\label{eq:stokes2}
\end{eqnarray}
The following boundary conditions are used:
\begin{eqnarray}
\vv&=&0 \quad  \quad \text{on the hot substrate,}\label{noslipsubstrate}\\
\vv&\to&0 \quad  \quad \text{far away from the drop,}\label{bcinfinity}\\
\vv \cdot \vtau &=&0  \quad \text{and}\quad \vv \cdot \vn =j \quad \text{on the droplet surface.}
\label{eq:v_interf}
\end{eqnarray}
\noindent Here $\vv$ and $p$ are the dimensionless velocity and pressure fields in the vapour (cf.~the scales in table~\ref{tab:scales}), and $\vtau$ is the unit tangential vector. 
In the first boundary condition (\ref{eq:v_interf}), a possible internal flow in the droplet is neglected relative to the velocity scale in the vapour, hence no slip. 
The last boundary condition (\ref{eq:v_interf}) contains the driving factor of the flow field, where the normal velocity at the droplet surface is determined by the evaporation flux (\ref{eq:evaporation_flux}), which is in turn determined by the temperature field obtained from the formulation (\ref{eq:lapl})--(\ref{eq:BCT3}) in \S~\ref{Tfield}.  Similarly to \S~\ref{Tfield}, this hydrodynamic part of the problem is also solved numerically using COMSOL Multiphysics.

The computation results for the vapour velocity fields are added into figure~\ref{fig2}a, mirroring the temperature fields and also displaying the streamlines. For large relative heights $\delta$, the streamlines remain straight near the droplet's surface, indicating that the flow is almost spherically symmetric there and only slightly disturbed by the substrate. Farther from the droplet, the streamlines are significantly bent due to the substrate presence. Higher velocity field values are attained for smaller $\delta$. This is not only due to a profound maximum in the evaporation flux due to the substrate proximity at small $\delta$ (as in figure~\ref{fig2}b1), but also additionally due to a confinement effect in a thin vapour layer between the droplet and the substrate, when the longitudinal velocity becomes even higher than the evaporation-flux-driven normal one at the droplet surface.  
 
\subsection{Levitation force}
\label{levitforce}
The bending and asymmetry of the evaporative flow due to the substrate gives rise to a hydrodynamic force acting on the evaporating droplet in the sense of its repulsion from the substrate. We refer to it as an evaporative force $F_{\rm{ev}}$. In our configuration (figure~\ref{fig1}), this amounts to a force acting on the droplet vertically upwards (along the $z$ axis), which is responsible for droplet levitation against gravity.  The force balance on the droplet and its levitation height are considered in \S~\ref{quasi-sta_lev} later on. Here we simply calculate $F_{\rm{ev}}$ in dimensionless terms (cf.~the scales in table~\ref{tab:scales}) as a function of the relative height $\delta$. Namely, we evaluate 
\begin{equation}
F_{\rm{ev}} = \left(\iint \limits_S (-p \vn +(\grad{\vv}+\grad{\vv}^{\intercal})\cdot \vn)\, \mathtt{d} S \right) \cdot \vz\, 
\label{eq:Fev}
\end{equation}
using the velocity and pressure fields computed in \S~\ref{sec:velocity}, where $\vz$ is a unit vector along $z$. 

\begin{figure}
  \centerline{\includegraphics[width=0.6\columnwidth]{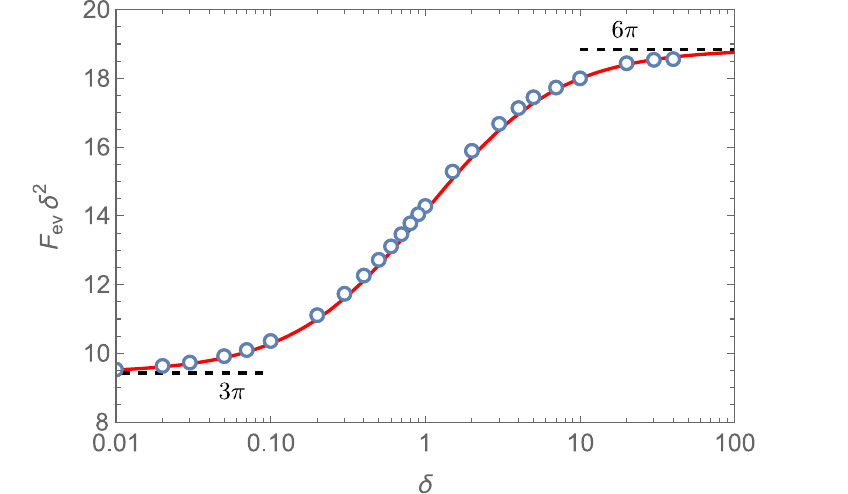}}
  \caption{Evaporative force $F_{\rm{ev}}$ (dimensionless, in terms of $\delta^2 F_{\rm{ev}}$) as a function of the relative droplet height $\delta$. Numerical results (blue open circles) are well fitted by equation~(\ref{eq:FitFevbis}) (red line). Black dashed lines correspond to the asymptotic behaviours.}
\label{fig:Fev}
\end{figure}

The result is reported in Fig.~\ref{fig:Fev} in terms of $F_{\rm{ev}} \delta^2$. The overall tendency is $F_{\rm{ev}}\propto\delta^{-2}$, as it is already known in the literature \citep{2012_Celestini,2012_Pomeau}. However, it is less known that the prefactor is different in the limits $\delta\to0$ and $\delta\to\infty$. For instance, \cite{2012_Celestini} attempted to fit the experimental data using a single prefactor.  We obtain a prefactor $3\pi$ as $\delta\to 0$, which can be deduced from the lubrication approximation \citep[cf.][see also Appendix~\ref{Fevasy0}]{2012_Pomeau,2021_Sobac_Erratum}, although \cite{2012_Pomeau} obtained $3\pi/8$ here (erroneously, in our opinion). 
In contrast, the prefactor is $6\pi$ as $\delta\to\infty$, which is confirmed by an asymptotic analysis described in Appendix~\ref{Fevasyinf}, where a number of conributions in terms of the droplet--substrate interaction are followed through. 
The following simple expression fits nicely our numerical result while respecting the prefactor values in both limits: 
\begin{equation}
 F_{\rm{ev}}(\delta)= \frac{3 \pi}{\delta^2}\, \frac{1+2\delta}{1+\delta} \,.
  \label{eq:FitFevbis}
\end{equation}
It covers the numerical data with a relative error of 1.4\%, whereas an even more precise fit is provided in Appendix~\ref{app:Fev}. 

\subsection{Validity of assumptions} 
\label{sec:limitations}

\begin{itemize}
  \item \textit{Negligible advection.} We start with the estimation of an evaporative P\'eclet number $\Pe=[\vv] R/\alpha_v$, where the evaporative velocity scale $[\vv]$ from table~\ref{tab:scales} is used. We obtain 
\begin{equation}
\Pe=\frac{c_{p,v} \Delta T}{\L} \,,
\label{Peclet}
\end{equation}
which incidentally turns out to be a version of the Jakob number often used in the literature. For the typical $\Delta T$ value (cf.~\S~\ref{formul}) and parameter values (cf.~the first raw of table~\ref{tab:properties} in Appendix~\ref{app:Properties}), we obtain $\Pe \approx 0.19\ll1$, which justifies the approximation used in \S~\ref{Tfield}.  
  \item \textit{Stokes approximation.} Likewise, the Reynolds number is $\Rere=\rho_v R [\vv] / \mu_v=\Pe \,\Prpr^{-1}$, where $\Prpr=\mu_v/(\rho_v\alpha_v)$ is the Prandtl number. As $\Prpr=0.71$ (cf.~ibid), we obtain $\Rere\approx 0.26\ll1$, confirming the approximation used in \S~\ref{sec:velocity}.
  \item \textit{Negligible natural convection.} This is related to small values of the Grashof number at the droplet scale 
\begin{equation}
\Gr=\frac{\rho_v g R^3}{\mu_v\nu_v}
\label{Gr}
\end{equation}
(written in this form given that the variations of $\rho_v$ are here of the order of $\rho_v$ itself). One typically obtains $\Gr<0.01$ for our small droplets ($R\lesssim 50~\mu\text{m}$). 
  \item \textit{Gas phase quasi-steadiness.} The results of the present section imply the quasi-steadiness of the temperature and velocity fields in the entire region between the droplet and the substrate, which may be especially questionable for large levitation heights $h$. The appropriate thermal and viscous time scales can be chosen as $\tau_{\rm{th}}=\max(R,h)^2/\alpha_v$ and $\tau_{\rm{vis}}=\rho_v \max(R,h)^2/\mu_v$. The quasi-steadiness takes place when $\tau_{\rm{th}}\ll\tau$ and $\tau_{\rm{vis}}\ll\tau$, where $\tau$ is the typical time scale of the process. As $\Prpr=O(1)$ here, we just limit our attention to the first one of these conditions for the sake of brevity, and hence 
\begin{equation}
\frac{\max(R,h)^2}{\alpha_v}\ll \tau\,.
\label{cond_gen}
\end{equation}
An immediately obvious time scale of the process is here the evaporation time scale of the droplet $\tau_{\rm{ev}}=\rho_l R^3/[J]$ (cf.~table~\ref{tab:scales} for $[J]$), i.e. 
\begin{equation} 
\tau_{\rm{ev}}=\frac{\rho_l \L R^2}{\lambda_v \Delta T}=\frac{\rho_l}{\rho_v}\frac{1}{\Pe}\frac{R^2}{\alpha_v} \,.
\label{tauev}
\end{equation}
Using it as $\tau=\tau_{\rm{ev}}$ in (\ref{cond_gen}), we arrive at 
\begin{equation}
\max(\delta^2,1)\ll \frac{\rho_l}{\rho_v} \frac{1}{\Pe} \,.
\label{cond_ev}
\end{equation}
Given that $\rho_l\gg\rho_v$ and $\Pe\ll 1$ here, the condition (\ref{cond_ev}) leaves a considerable margin for possible large values of the relative height $\delta=h/R$. We shall come back to it later on, after having considered concrete solutions for $h$. 
\end{itemize}

\section{Levitation and take-off: quasi-steady approach}
\label{quasi-sta_lev}

\subsection{Quasi-steady approach as such}
\label{quasista}

\begin{figure}
  \centerline{\includegraphics[width=\columnwidth]{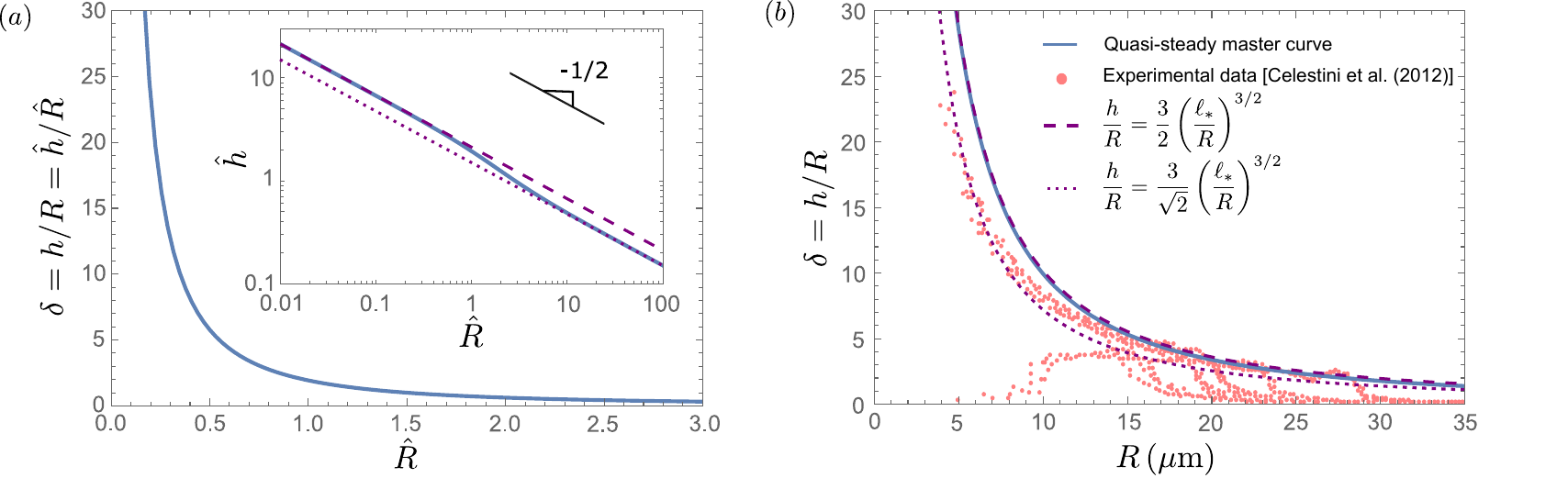}}
  \caption{(a) Relative height as a function of the droplet radius as predicted by the quasi-steady model. Full solution (solid blue line, `master curve') and the asymptotic behaviours for smaller and larger radiuses (dotted and dashed purple lines, respectively). The log-log inset highlights the dominant power law and the prefactor change between the two limits. (b)~First comparison with the experimental data of \citet{2012_Celestini} for water (with $\Delta T=300^{\circ}$C).}
\label{fig:QSMasterCurve}
\end{figure}

A vertical balance of the (evaporative) levitation force (\ref{eq:FitFevbis}), where recall its scale in table~\ref{tab:scales} and the definition (\ref{ratio}), against the droplet weight directly yields the equation for the levitation height of the droplet $h$ as a function of its radius $R$:
\begin{equation}
3\pi \frac{\mu_v\lambda_v \Delta T}{\rho_v \mathcal{L}} \frac{R^2}{h^2}\, \frac{R+2 h}{R+h}=\frac{4\pi}{3}\rho_l g R^3 \,.
\label{eqQSdim}
\end{equation}
There exists a single natural length scale $\ell_*$, the non-dimensionalization with which renders equation (\ref{eqQSdim})  parameter-free: 
\begin{equation}
\frac{1}{\hat{h}^2}\, \frac{\hat{R}+2 \hat{h}}{\hat{R}+\hat{h}}=\frac{4}{9} \hat{R}\, ,
\label{eqQSadim1}
\end{equation}
where
\begin{equation}
[\hat{R}]=[\hat{h}]=\left(\frac{\mu_v\lambda_v \Delta T}{\rho_v \rho_l g \mathcal{L}}\right)^{1/3}\equiv\ell_*\,,\quad \hat{R}=\frac{R}{[\hat{R}]}\,,\ \hat{h}=\frac{h}{[\hat{h}]}\, .
\label{scales1}
\end{equation}
The solution of (\ref{eqQSadim1}) is shown in figure~\ref{fig:QSMasterCurve}(a) and adheres to the following asymptotic behaviour: 
\begin{eqnarray}
\label{eq:h}
\hat{h}&=&\frac{3}{2}\frac{1}{{\hat{R}^{1/2}}}\, \, \, \, \, \,   \text{i.e.}\, \, \delta=\frac{3}{2}\frac{1}{{\hat{R}^{3/2}}}\, \, \, \, \, \,  \text{as}\, \hat{R}\to\infty \,,
\\
\hat{h}&=&\frac{3}{\sqrt{2}}\frac{1}{{\hat{R}^{1/2}}}\, \text{i.e.}\, \, \delta=\frac{3}{\sqrt{2}}\frac{1}{{\hat{R}^{3/2}}}\, \text{as}\, \hat{R}\to0 \,. 
\label{eq:h2}
\end{eqnarray}
The length scale $\ell_*$ indicates the characteristic size $R\sim\ell_*$ at which the droplet takes off at a height of the order of itself, with $h \sim R$ (i.e.~$\delta\sim 1$). At smaller sizes ($R\ll \ell_*$), the droplet soars even higher ($h\gg\ell_*\gg R$, $\delta\gg 1$), whereas at larger sizes ($R\gg \ell_*$), the droplet levitates lower ($h\ll\ell_*\ll R$, $\delta\ll 1$). Typically, $\ell_*$ is in the range of a few tens of micrometers. For our reference case of a water droplet on a superheated substrate with $\Delta T=300^{\circ}$C, we obtain $\ell_*=28.46~\mu$m, which is much smaller than the capillary length $\ell_c=\sqrt{\gamma/(\rho_l g)}$ ($\gamma$ being the liquid--air surface tension, $\ell_c\sim 2.5\,\mathrm{mm}$ for water). This `take-off scale' $\ell_*$ has earlier been pointed out by \cite{2012_Celestini} and \cite{2012_Pomeau} (their notation $R_l$) as the scale at which a \textit{drastic} take-off takes place \citep[although note that a mere increase of $h$ as $R$ decreases already starts from much larger sizes $R$, cf.][as well as~\S\ref{global} here]{2021_Sobac_Erratum}. They also interpret it as the droplet size starting from and below which the lubrication approximation in the vapour film between the droplet and the substrate becomes invalid (since $h\ll R$ ceases indeed to hold). Note that $h=R$ for $R=3/2\, \ell_*$.

Similarly to what was commented for $F_{\rm{ev}}$ in \S\ref{levitforce}, the overall tendency $h\sim R^{-1/2}$ (or $h/R\sim R^{-3/2}$) is well-known since \cite{2012_Celestini} and \cite{2012_Pomeau}, who first pointed out this exponent. However, we here calculate the prefactor and point out that it is actually not fully constant, as hightlighted in the inset of figure~\ref{fig:QSMasterCurve}(a) and further put into evidence by the two different limiting values in  (\ref{eq:h}) and (\ref{eq:h2}). 

A first comparison with experiment is undertaken in figure~\ref{fig:QSMasterCurve}(b), where we consider just the upper layer of experimental points, while the points lower than that are deemed to belong to some transients (see also a remark in \S\ref{dyn_gen} later). 
It is important to recall that these experiments dealt with small droplets, typically ranging in radius from $1\,\mu\mathrm{m}$ to $30\,\mu\mathrm{m}$, which is of the order of or smaller than $\ell_*$ here. For this size range ($\hat{R}\lesssim 1$), specific to take-off observations, the full solution of (\ref{eqQSadim1}) already practically coincides with the asymptotic limit (\ref{eq:h2}), cf.~figure~\ref{fig:QSMasterCurve}. 
While this seems to agree well with experiment for $15\,\mu\mathrm{m}\lesssim R\lesssim 30\,\mu\mathrm{m}$, an overprediction is nonetheless observed as the droplet size decreases below $R\lesssim 15\,\mu\mathrm{m}$.  
Astonishingly, we observe that it is rather the asymptotic behavior (\ref{eq:h}) that starts to get closer to the experimental points.  
However, this must be by chance or for a wrong reason, since equation (\ref{eq:h}), with the prefactor it contains, is appropriate in the limit of larger droplets (and not the smaller ones we are discussing right now). We shall come later to what 
will be the right explanation here. 

The droplet radius $R$ decreases over time by evaporation (rather than just being a given constant parameter), and hence $h$, related to $R$ by (\ref{eqQSdim}), is also a function of time. The steady force balance (\ref{eqQSdim}) or (\ref{eqQSadim1}) is then assumed to be valid in a quasi-steady sense, and $R(t)$ and $h(t)$ follow the solid curve of Fig.~\ref{fig:QSMasterCurve} as time goes on.  
The mass of the droplet $\frac{4\pi}{3} \rho_l R^3$ decreases in time at an evaporation rate  given by (\ref{eq:FitJbis}) with the scale from table~\ref{tab:scales}. This balance gives rise to the following equation for the droplet radius evolution:
\begin{equation}
\rho_l R\, \frac{\ddd R}{\ddd t}=-\frac{\lambda_v \Delta T}{\mathcal{L}} \left[ 1+\frac{1}{2} \ln\left(1+\frac{R}{h}\right) \right] \,.
\label{eq:R}
\end{equation}
Using the time scale 
\begin{equation}
[\hat{t}]=\frac{\rho_l \mathcal{L} {\ell_*}^2}{\lambda_v \Delta T}=\frac{\rho_l}{\rho_v}\frac{1}{\Pe}\frac{\ell_*^2}{\alpha_v}\equiv\tau_*\, 
\label{eq:scale1_time}
\end{equation}
alongside the scales (\ref{scales1}), equation (\ref{eq:R}) is rendered free of any parameters similarly to (\ref{eqQSadim1}). Namely, we arrive at  
\begin{equation}
\hat{R}\, \frac{\ddd \hat{R}}{\ddd \hat{t}}= -1-\frac{1}{2} \ln\left(1+\frac{\hat{R}}{\hat{h}}\right) \,. 
\label{eq:Radim}
\end{equation}

\begin{figure}
  \centerline{\includegraphics[width=\columnwidth]{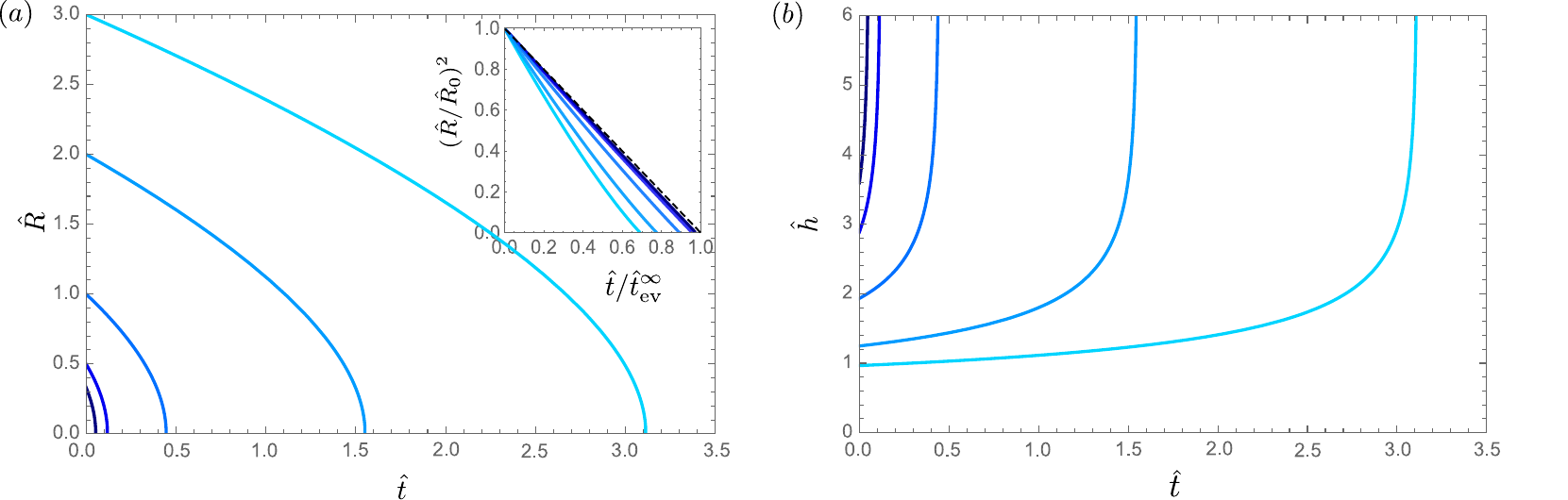}}
  \caption{Dimensionless evolutions of the radius $\hat{R}$ and the height  $\hat{h}$ of a spherical Leidenfrost droplet over time $\hat{t}$ computed by the quasi-steady model (coupled equations \ref{eqQSadim1} and \ref{eq:R}) for initial droplet radii $\hat{R}_0=\{1/3,1/2,1,2,3\}$ (from dark to light blue). The corresponding initial quasi-steady heights are $\hat{h}_0=\{3.60, 2.89, 1.93, 1.25, 0.97\}$. The inset serves to illustrate the extents to which the $R^2$-law holds and to which the evaporation is accelerated by the superheated substrate, where the time is normalized to the evaporation time of a freely suspended droplet.}
\label{fig:QhR(t)}
\end{figure}

Now the dimensionless evolution problem for $\hat{R}(\hat{t})$ and $\hat{h}(\hat{t})$ is defined by a system of two coupled equations (\ref{eqQSadim1}) and (\ref{eq:Radim}), for which the initial conditions $\hat{R}=\hat{R}_0$ and $\hat{h}=\hat{h}_0$ at $\hat{t}=0$ are posed with $\hat{R}_0$ and $\hat{h}_0$ not being independent but rather related by (\ref{eqQSadim1}). Figure \ref{fig:QhR(t)} illustrates the (numerically obtained) solution for various initial droplet radii $\hat{R}_0=\{1/3,1/2,1,2,3\}$. Evidently, the curves demonstrate that $\hat{R}$ decreases over time due to evaporation until extinction, with larger droplets exhibiting longer lifespans. Concurrently, $\hat{h}$ increases over time as the droplet size decreases, larger droplets being closer to the substrate at the initial time in accordance with (\ref{eqQSadim1}). It is important to note that, within the present quasi-steady description, a Leidenfrost droplet takes off reaching an infinite height $\hat{h}\to\infty$ at the end of its life (as $\hat{R}\to0$), in accordance with (\ref{eq:h2}). 

In the inset, $(\hat{R}/\hat{R}_0)^2$ is plotted as a function of $\hat{t}/\hat{t}_{\rm{ev}}^{\infty}$ in order to highlight the evaporative behavior of a spherical Leidenfrost droplet as compared to the well-known limit case of a spherical droplet suspended in an unbounded gas medium. Here $\hat{t}_{\rm{ev}}^{\infty}=(\hat{R}_0)^2/2$ is the dimensionless evaporation time of such a suspended droplet (which can be derived from equation~(\ref{eq:Radim}) in the limit $\delta=\hat{h}/\hat{R}\to\infty$).
Owing to the interaction with the superheated substrate, appearing through the logarithmic term in eq.~(\ref{eq:Radim}), the well-known $R^2$-law is recovered only for large values of $\hat{h}_0$. Thus, $\hat{R}^2$ generally does not linearly decrease in time, while these droplets evaporate faster than their suspended counterparts due to the proximity of the superheated substrate. 

Needless to note that by parametrically plotting $\hat{h}(\hat{t})/\hat{R}(\hat{t})$ or $\hat{h}(\hat{t})$ as a function of $\hat{R}(\hat{t})$ ($t$ being the parameter) we retrieve the same `master' curve as depicted in Fig.~\ref{fig:QSMasterCurve}.
Within the present quasi-steady approach, such a master curve is just trivially given by an algebraic equation like (\ref{eqQSdim}) or (\ref{eqQSadim1}). However, this may become less trivial in what follows. 

\subsection{Validity of assumptions} 
\label{sec:limitations2}

The quasi-steady result that the droplet soars to an infinite height at the end of its life, cf. equation (\ref{eq:h2}) and figure~\ref{fig:QhR(t)}~(right), looks suspicious from the physical point of view. One can wonder whether the quasi-steadiness criterium (\ref{cond_ev}) for the fields in the gas phase is still fulfilled in view of $\delta\to\infty$. Furthermore, $h\to\infty$ also implies infinite velocity and acceleration  ($\mathtt{d}h/\mathtt{d}t\to \infty$ and $\mathtt{d}^2 h/\mathtt{d}t^2\to \infty$), and hence one can wonder whether the steady force balance such as (\ref{eqQSdim}) is still adequate with neglecting the drag (proportional to $\mathtt{d}h/\mathtt{d}t$) and inertia (proportional to $\mathtt{d}^2 h/\mathtt{d}t^2$) forces on the droplet.  Below, we make an estimation of those effects when the only source of unsteadiness is the evaporation of the droplet, i.e.~when the time scale is $\tau=\tau_{\rm{ev}}$ as given by (\ref{tauev}).  

Disregarding numerical prefactors, the inertia, (Stokes) drag and levitation forces can be estimated as 
\begin{equation}
\mathrm{inertia}\sim \rho_l R^3 \frac{h}{\tau_{\rm{ev}}^2}\,,\quad 
\mathrm{drag}\sim \mu_v R \frac{h}{\tau_{\rm{ev}}}\,,\quad \mathrm{levitation}\sim \frac{\mu_v\lambda_v \Delta T}{\rho_v \mathcal{L}} \frac{R^2}{h^2}\,,
\label{force_estims}
\end{equation}
where the estimation of the levitation force is just based on the left-hand side of (\ref{eqQSdim}). Using the expression (\ref{tauev}) for $\tau_{\rm{ev}}$, one can immediately see that 
\begin{equation}
\frac{\mathrm{inertia}}{\mathrm{drag}}\sim \Prpr^{-1} \Pe
\label{inertia_to_drag}
\end{equation}
As we have $\Prpr\sim 1$, $\Pe\ll 1$ here (cf.~\S\ref{sec:limitations}), inertia can be disregarded against the drag in the present context with $\tau=\tau_{\rm{ev}}$ (which does not exclude that intertia can be essential in other contexts, cf.~\S\ref{dyn_gen} later on).  Then, it just remains to compare the drag and levitation forces. Using (\ref{force_estims}) on account of (\ref{tauev}), one can obtain 
\begin{equation}
\frac{\mathrm{drag}}{\mathrm{levitation}}\sim \frac{\rho_v}{\rho_l} \left(\frac{h}{R}\right)^3\,.
\label{drag_to_levitation}
\end{equation}
Thus, the drag can be neglected in favour of a quasi-steady force balance like (\ref{eqQSdim}) provided that 
\begin{equation}
\delta^3\ll \frac{\rho_l}{\rho_v}\,.
\label{cond_QSforce}
\end{equation}
Given that the liquid density is much greater then the vapour density ($\rho_l/\rho_v\gg 1$), the condition (\ref{cond_QSforce}) leaves quite a considerable margin for the present quasi-steady approach to be valid. It is only for sufficiently small droplets levitating too high (such that $\delta\sim (\rho_l/\rho_v)^{1/3}$) that  it breaks down and the drag force should be incorporated (but still not the inertia force, according to the earlier estimations), as intuitively expected. Moreover, one can clearly see that the condition (\ref{cond_QSforce}) is more restrictive in the realm of large $\delta$ than (\ref{cond_ev}), which is further reinforced by the fact that $\Pe\ll 1$. This means that even when the drag force becomes important, the temperature and velocity fields between the droplet and the substrate can still be regarded quasi-steady, and hence the expressions like (\ref{eq:FitJbis}) and (\ref{eq:FitFevbis}) are still valid. An analysis aiming at smaller $R$ (and larger $h$) and incorporating the drag force is realized in \S\ref{basics2} and \S\ref{dropdyn} below.  
 
In the opposite limit of larger $R$ (and smaller $h$), the validity of the quasi-steady approach as used here is therefore not put into question. However, it is rather the full-sphericity assumption that becomes more restrictive, when (even still within $R\ll \ell_c$ and a practical sphericity of the most of the droplet) a small part of the droplet bottom gets flattened by gravity \citep{2012_Pomeau} with an essential effect from the Leidenfrost viewpoint. 
In this regard, the result (\ref{eq:h}) should be understood in an intermediate asymptotic sense, as valid for $R\gg \ell_*$ but $R$ still much smaller than the bottom-flattening scale. This will be considered in more detail in \S\ref{global}.

\section{Basic calculations (continued): drag force}
\label{basics2}

As stipulated in \S\ref{sec:limitations2}, the drag force, $F_\text{drag}$, is required for further analysis. The present section, dedicated to $F_\text{drag}$, is organized as a continuation of \S\ref{basics} and mirrors the same style as far as notations, non-dimensionalization and scales are concerned. The scales relevant here are summarized in table~\ref{tab:scales2} (which is the counterpart of table~\ref{tab:scales} there), where $U$ is the droplet (translation) velocity in the vertical direction (its only component here).

\begin{table*}
\centering\begin{tabular}{ccccc}
  \hline
  $[r,z,s]$ & $\quad [S]$ & $\quad [\vv]$ & $\quad [p]$  & $\quad [F_\text{drag}]$ \\
  \hline
 $R$ & $\quad R^2$ & $\quad U$ & $\quad\mu_v \frac{U}{R}$ & $\quad-\mu_v R \, U$ \\ 
  \hline
  \end{tabular}
\caption{Scales used to render the various quantities dimensionless in \S\ref{basics2}. The square brackets denote the scale of the quantity inside.}
\label{tab:scales2}
   \end{table*}

As the primary need for such a consideration arose in the context of large $\delta$ (cf.~\S\ref{sec:limitations2}), a mere use of the (dimensionless) Stokes drag $F_\text{drag}=6\pi$ in an unbounded medium could be quite sufficient here (as  well as in \S\ref{dropdyn} that follows), where the rigid-sphere prefactor $6\pi$ is used on account of the liquid dynamic viscosity being much larger than the vapour one. Nonetheless, for the sake of generality, we shall here proceed implying $\delta=O(1)$, all the more so that it will be particularly relevant later on in the context of \S\ref{dyn_gen}. Thus, the goal is to compute $F_\text{drag}(\delta)$. 

To this purpose, we once again solve the dimensionless Stokes equations (\ref{eq:stokes1}) --(\ref{eq:stokes2}) with the boundary conditions (\ref{noslipsubstrate})--(\ref{bcinfinity}) (although the dimensional scales are now different and given by table~\ref{tab:scales2}). However, the `evaporation' boundary conditions (\ref{eq:v_interf}) are now replaced with 
\begin{equation}
\vz \cdot \vv=1
\label{eq/v_interf2}
\end{equation}
reflecting droplet translation along $z$. Finally, the same expression as on the right-hand side of (\ref{eq:Fev}) is used to compute $F_\text{drag}$.

\begin{figure}
  \centerline{\includegraphics[width=0.6\columnwidth]{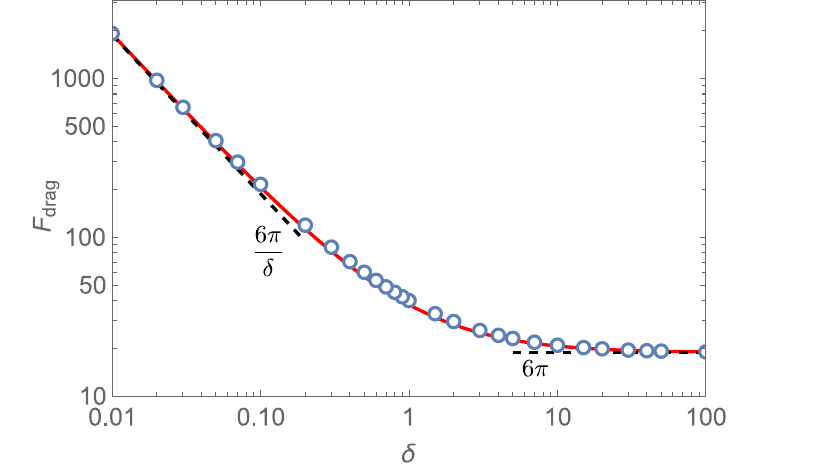}}
  \caption{Dimensionless drag force as a function of the relative droplet height $\delta$. Numerical results (blue open circles) well-agree with the classical approximation~\ref{eq:FitFdragbis} (red line). Black dashed lines correspond to the asymptotic behaviours.}
\label{fig:Fdrag}
\end{figure}

The computation results are illustrated in Fig~\ref{fig:Fdrag} together with the following approximate expression~\citep{guyon2012hydrodynamique}: 
 \begin{equation}
F_{\rm{drag}}(\delta)= 6 \pi \left(1 + \frac{1}{\delta} \right)
 \label{eq:FitFdragbis}
 \end{equation}
While strictly valid in the limits $\delta\to0$ and $\delta\to\infty$, the result (\ref{eq:FitFdragbis}) can be seen to deviate from the numerical results up to 7$\%$ for intermediate values of $\delta\sim1$. A more precise fit of the numerical data is proposed in Appendix~\ref{app:DragForce}. Nonetheless, we shall stick in the present study to a simpler and elegant expression (\ref{eq:FitFdragbis}), which will be sufficient for our purposes. 

\section{Levitation and take-off: (indispensable) dynamic approach for smaller droplets}
\label{dropdyn}

We now turn to the case of smaller droplets (levitating higher), for which the quasi-steady approach followed through in \S\ref{quasista} breaks down. As pointed out in \S\ref{sec:limitations2}, the drag force, depending on the droplet translation velocity as it soars higher while vaporizing, becomes essential (although not the inertia force). Supplementing the quasi-steady force balance (\ref{eqQSdim}) with the drag force, we obtain 
\begin{equation}
-6\pi\mu_v R \left(1+\frac{R}{h}\right) \frac{\ddd h}{\ddd t} +3\pi \frac{\mu_v\lambda_v \Delta T}{\rho_v \mathcal{L}} \frac{R^2}{h^2}\, \frac{R+2 h}{R+h}=\frac{4\pi}{3}\rho_l g R^3 \,. 
\label{eqdyndim}
\end{equation}
Here the dimensionless expression (\ref{eq:FitFdragbis}) multiplied by the scale from table~\ref{tab:scales2} has been used with $U=\frac{\ddd h}{\ddd t}$. (Strictly speaking, the velocity of the center of mass is rather given by $\frac{\ddd h}{\ddd t}+\frac{\ddd R}{\ddd t}$, but we shall be disregarding such a nuance here.)

The natural distinguished scales are such that all terms in (\ref{eqdyndim}) and in (\ref{eq:R}) are of the same order of magnitude:
\begin{eqnarray}
[\tilde{R}]=\left(\frac{\rho_v}{\rho_l}\right)^{2/9}\ell_*\,,\quad [\tilde{h}]=\left(\frac{\rho_v}{\rho_l}\right)^{-1/9}\ell_*\,,\quad [\tilde{t}]=\left(\frac{\rho_v}{\rho_l}\right)^{4/9} \tau_*\,, 
\label{scales3} \\
\tilde{R}=\frac{R}{[\tilde{R}]}\,,\quad \tilde{h}=\frac{h}{[\tilde{h}]}\,,\quad \tilde{t}=\frac{t}{[\tilde{t}]}\,,\quad \epsilon\equiv\frac{[\tilde{R}]}{[\tilde{h}]}=\left(\frac{\rho_v}{\rho_l}\right)^{1/3} \,,
\label{vars_tilde}
\end{eqnarray}
and the dynamical system becomes 
\begin{equation}
- \left(1+\epsilon\frac{\tilde{R}}{\tilde{h}}\right) \frac{\ddd \tilde{h}}{\ddd\tilde{t}} +\frac{\tilde{R}}{\tilde{h}^2}\, \frac{\tilde{h}+\frac{1}{2}\epsilon \tilde{R}}{\tilde{h}+\epsilon\tilde{R}}=\frac{2}{9} \tilde{R}^2 \,,
\label{eqdynadim}
\end{equation}
\begin{equation}
\tilde{R}\,\frac{\ddd \tilde{R}}{\ddd \tilde{t}}=-1-\frac{1}{2} \ln\left(1+\epsilon\frac{\tilde{R}}{\tilde{h}}\right) \,.
\label{eq:Radim2}
\end{equation}
The distinction between the present scales (\ref{scales3}) and the previously considered ones (\ref{scales1}) and (\ref{eq:scale1_time}) is eventually owing to a small parameter given by the vapour-to-liquid density ratio $\rho_v/\rho_l\ll 1$. 
We note that the scales $[\tilde{R}]$ and $[\tilde{h}]$ are different here, and the typical relative height of the droplet levitation is now $\delta\sim (\rho_l/\rho_v)^{1/3}\gg1$. It is exactly at such values of~$\delta$ that the criterion (\ref{cond_QSforce}) of the quasi-steady force balance breaks down, as expected, which confirms the coherence of the present dynamic approach.  
At the same time, the criterion~(\ref{cond_ev}) of the quasi-steadiness of the temperature and velocity fields between the droplet and the substrate is still satisfied, as already discussed in \S\ref{sec:limitations2}. 

\begin{figure}
  \centerline{\includegraphics[width=\columnwidth]{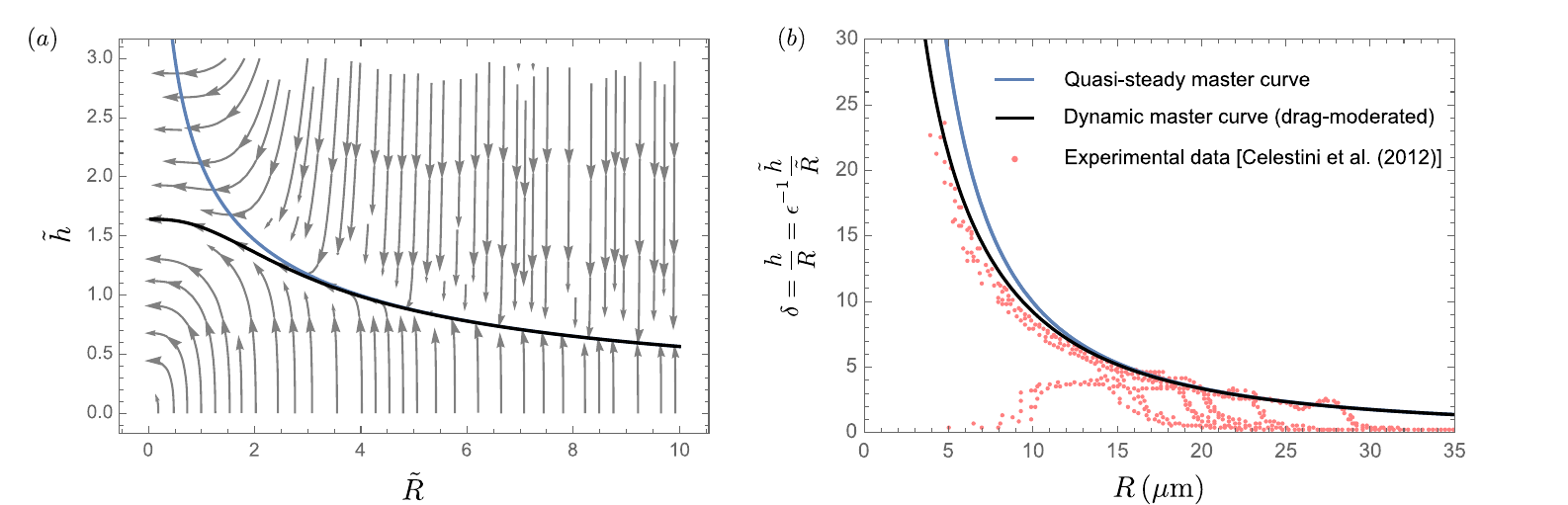}}
  \caption{(a) Phase portrait of the dynamical system for $\tilde{R}(\tilde t)$ and $\tilde{h}(\tilde t)$ appropriate in the limit of smaller droplet levitating higher (incorporating the drag force upon the quasi-steady force balance) and in the scaling appropriate to that limit (tilded variables). Parameter values used: $\epsilon=0.076$, corresponding to water droplet experiments by \cite{2012_Celestini} at $\Delta T=300^\circ\mathrm{C}$ (cf.~Appendix~\ref{app:Properties}). The blue solid line shows the earlier obtained quasi-steady `master curve', which coincides with the separatrix (black solid line, `dynamic master curve') for larger $\tilde{R}$ but diverges ($\tilde{h}\to\infty$) for smaller $\tilde{R}$. (b)~Comparison with the mentioned experiment using the quasi-steady and dynamic master curves.}
\label{fig:sysdyn}
\end{figure}

The phase portrait of the dynamical system (\ref{eqdynadim})--(\ref{eq:Radim2}) is represented in figure~\ref{fig:sysdyn}(a) (generated using the \texttt{StreamPlot} command in \textsl{Mathematica}). 
Notably, while a Leidenfrost droplet is theoretically expected to ascend indefinitely within the quasi-steady approach, the presence of the drag force results in a saturation of $\tilde{h}$ towards a finite take-off value. In other words, the Leidenfrost droplets always vanish ($\tilde{R}\to 0$) at a finite height. Physically, the drag force, represented by the first term on the left-hand side of (\ref{eqdyndim}), becomes so significant towards the end of life of the droplet that it blocks the soaring tendency dictated by the levitation force (the second term ibid). In principle, the droplet can end up at whatever height, depending on the initial conditions (figure~\ref{fig:sysdyn}a). However, there is a distinguished value of the final levitation height valid for most of the droplets, at least for those starting off from a sufficiently large size. In such a case, the phase trajectory is seen to fastly reach the separatrix (figure~\ref{fig:sysdyn}a), along which the droplet evolution ensues until the droplet vanishes at the distinguished final height corresponding to the separatrix: 
\begin{equation}
\tilde{h}_\text{fin}=1.69 (1 + 0.35\, \epsilon)\,,
\label{hfin_adim}  
\end{equation}
which was computed numerically assuming a linear dependence on the small parameter $\epsilon$. Using the scales (\ref{scales3}) on account of (\ref{scales1}), this can be rewritten in dimensional terms: 
\begin{equation}
h_\text{fin}=1.69  \left(\frac{\rho_l}{\rho_v}\right)^{1/9} \left(\frac{\mu_v\lambda_v \Delta T}{\rho_v \rho_l g \mathcal{L}}\right)^{1/3} \left(1 + 0.35\left(\frac{\rho_v}{\rho_l}\right)^{1/3} \right) \,.
\label{hfin_dim}  
\end{equation}
For instance, under the conditions of the experiments by~\cite{2012_Celestini} with water droplets (cf.~Appendix~\ref{app:Properties} for the parameters), we obtain $\tilde{h}_\text{fin} = 1.64$ and $h_\text{fin}=110.36\,\mu\mathrm{m}$. The separatrix now becomes our new, dynamic `master curve'. It replaces the quasi-steady one, soaring to infinity ($\tilde{h}\to\infty$) as $\tilde{R}\to 0$ and corresponding to
\begin{equation}
\frac{1}{\tilde{h}^2}\, \frac{\tilde{h}+\frac{1}{2}\epsilon\tilde{R}}{\tilde{h}+\epsilon\tilde{R}}=\frac{2}{9} \tilde{R}\, 
\label{eqQSadim2}
\end{equation}
in terms of the tilded variables, which is also depicted in figure~\ref{fig:sysdyn} (solid black line) for comparison. The dynamic master curve is different for smaller droplets ($\tilde{R}\lesssim 3$), 
whereas for larger droplets the quasi-steady result is recovered (the solid black curve coinciding with the separatrix in figure~\ref{fig:sysdyn}a), as expected.

Figure~\ref{fig:sysdyn}(b) undertakes a direct comparison with the experiments by \cite{2012_Celestini} for water droplets at $\Delta T=300^\circ\mathrm{C}$. We see that the dynamic model, considered in the present section, shows a noticeable improvement over the previously considered quasi-steady approach. The improvement is just manifest for smaller droplets ($R\lesssim 15\,\mu\mathrm{m}$).   

\section{Dynamic approach in general}
\label{dyn_gen}
 
The dynamic approach considered in \S\ref{dropdyn} came as indispensable and, in this sense, forced in the domain of smaller droplets, where the consideration of \S\ref{quasi-sta_lev} based on the quasi-steady force balance could no longer be valid. At the same time, this made it limited to that domain, where, in particular, only the drag force was essential while inertia could be disregarded. In the present section, we take up a full dynamic approach, which would presumably be valid in the entire range of droplet sizes considered in the present paper. In \S\ref{quasi-sta_lev}, we considered just the (quasi-)equilibrium positions (heights) of the droplet. Here, we inquire what the droplet behaviour will be if initially out of that equilibrium.  

Complementing the force balance (\ref{eqdyndim}) with intertia, we arrive at  
\begin{equation}
\frac{4\pi}{3} \rho_l R^3  \frac{\ddd^2 h}{\ddd t^2} =-6\pi\mu_v R \left(1+\frac{R}{h}\right) \frac{\ddd h}{\ddd t} +3\pi \frac{\mu_v\lambda_v \Delta T}{\rho_v \mathcal{L}} \frac{R^2}{h^2}\, \frac{R+2 h}{R+h}-\frac{4\pi}{3}\rho_l g R^3 \,.
\label{eqdyndim2}
\end{equation}
Thus, we end up with a third-order system of ODEs given by equations (\ref{eqdyndim2}) and (\ref{eq:R}). It can in principle be (numerically) solved starting from any initial condition $R=R_0$, $h=h_0$, $\frac{\ddd h}{\ddd t}=h'_0$ at $t=0$ to obtain $R(t)$ and $h(t)$, where $R_0>0$, $h_0>0$ and $h'_0$ are some initial values.

 \begin{figure}
  \centerline{\includegraphics[width=\columnwidth]{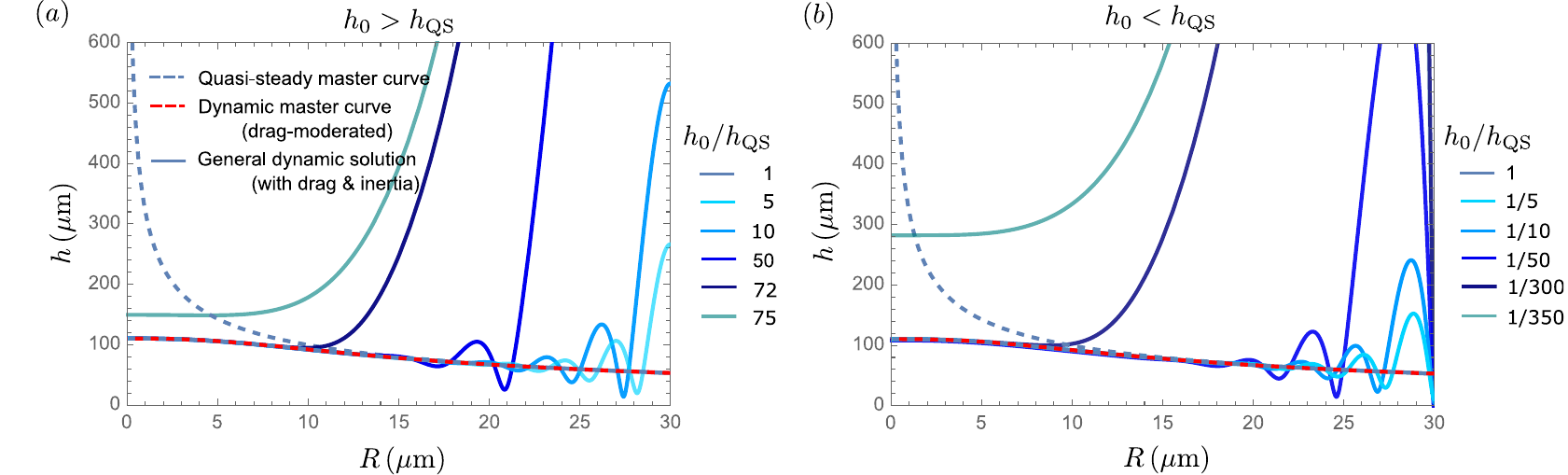}}
  \caption{Leidenfrost water droplet trajectories in the plane $h$ versus $R$ computed by means of the general dynamic approach (including drag and inertia) using the parameters of \cite{2012_Celestini} (cf.~Appendix~\ref{app:Properties}). Initial radius $R_0=30~\mu$m. Various initial heights $h_0$ are tested, which are here represented relative to the equilibrium quasi-steady height $h_{\rm{QS}}=50.24\,\mu\mathrm{m}$. The quasi-steady and dynamic master curves (blue and red dashed lines) are also shown for reference.}
\label{fig:Inertie}
\end{figure}

The solution is illustrated in figure~\ref{fig:Inertie} for a water Leidenfrost droplet of an initial radius $R_0=30~\mu$m starting from various initial heights $h_0$ with $h'_0=0$. As earlier, the parameters of the experiment by \cite{2012_Celestini} are used (i.e.~$\Delta T=300^{\circ}$C and 1~atm, cf.~Appendix~\ref{app:Properties}). We note that the chosen value of the initial radius is large enough for the quasi-steady approach to work well (cf.~figure~\ref{fig:sysdyn}b) and the corresponding quasi-steady height, obtained from equation (\ref{eqQSdim}) at $R=R_0$, is $h_{\rm{QS}}=50.24\,\mu\mathrm{m}$. The several initial heights tested are then conveniently expressed in the units of $h_{\rm{QS}}$. The previously obtained quasi-steady (\S\ref{quasi-sta_lev}) and dynamic (\S\ref{dropdyn}) master curves are also shown in figure~\ref{fig:Inertie} for reference. 

For $h_0=h_{\rm{QS}}$, we see that the solution adheres to the dynamic master curve, which coincides with the quasi-steady one for larger $R$ and is drag-force-moderated for smaller $R$, and where the incorporation of inertia into the force balance has practically no effect, as expected. For an initial height $h_0$ out of the equilibrium position $h_{\rm{QS}}$, the droplet aproaches the dynamic master curve relatively fast and in an oscillatory way (like a damped oscillator), and then the evolution continues along that curve. The droplets intially located too close to the substrate can rebound to considerable heights as propelled by the levitation force. The droplets starting from or propelled to considerable heights rejoin the dynamic master curve at a later time and smaller size. In this case, the rejoining already happens in a monotonic way, quite in accordance with the scenario for smaller droplets described in \S\ref{dropdyn}, where inertia could be disregarded. Furthermore, the droplets finding themselves at a certain moment excessively high vaporize at some finite height before reaching the dynamic master curve, which also forms part of that scenario. Under the conditions explored in figure~\ref{fig:Inertie}, the latter scenario occurs whenever $h_0\lesssim (1/300)h_{\rm{QS}}(R_0)=0.18\,\mu\mathrm{m}$ and $h_0\gtrsim 72 h_{\rm{QS}}(R_0)=3.83\,\mathrm{mm}$.

\begin{figure}
  \centerline{\includegraphics[width=0.55\columnwidth]{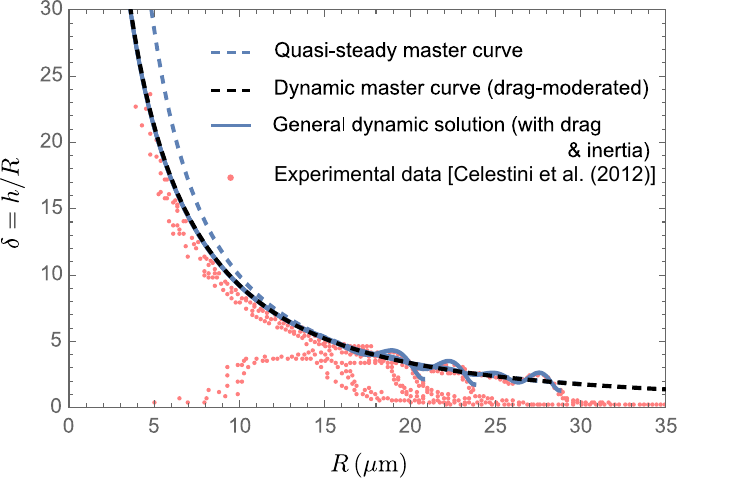}}
  \caption{Relative height $\delta=h/R$ of a water Leidenfrost droplet as a function of its radius $R$. The experimental data from \cite{2012_Celestini} are compared to the quasi-steady and dynamic (drag-moderated) master curves, as well as with the trajectories computed from the general dynamic approach  (including both drag and inertia) using the initial conditions $h_0/h_{\rm{QS}}(R_0)=3/2$, $R_0=\{20.8; \ 23.8; \ 28.8\}~\mu$m and $h'_0=0$.}
\label{fig:Inertie_Celestini}
\end{figure}

Small rebounds around the master curve can actually reproduce certain oscillatory trends observed in the experimental points by \cite{2012_Celestini}, as illustrated in figure~\ref{fig:Inertie_Celestini}. However, the nature of the experimental points located too close to the substrate remains unclear, the understanding of which may require staging further experiments and thinking of physical factors not included in the present model. The present modelling indicates that there may be a certain dependence on the manner the Leidenfrost droplets are deposited in experiment as the initial conditions can be such that the droplet fully evaporates before reaching the master curve. 

The consistency of the approach in regard of the oscillatory relaxation obtained here can be assessed as follows. Focusing just on larger droplets with $h\sim R$, the oscillation time scale $\tau_\text{oscil}=\sqrt{R/g}$ can be compared with the viscous and thermal time scales $\tau_\text{vis}\sim\rho_v R^2/\mu_v$ and $\tau_\text{th}\sim R^2/\alpha_v$ (the latter two are of the same order on account of $\Prpr\sim 1$ and can thus be used interchangeably in estimations). As $\tau_\text{vis}\tau_\text{th}/\tau_\text{oscil}^2=\Gr$, cf.~equation~(\ref{Gr}), while it was estimated $\Gr\ll 1$ in \S\ref{sec:limitations}, we see that  $\tau_\text{vis}\,,\,\tau_\text{th}\ll \tau_\text{oscil}$. Thus, the implied quasi-steadiness of the temperature and velocity fields does hold during the oscillation cycle, hence the sought consistency. 
Similarly, the ratio of the intertia and drag forces in (\ref{eqdyndim2}) can be estimated at $\sim\frac{\rho_l}{\rho_v}\Gr^{1/2}$ taking $\tau_\text{oscil}$ as the time scale. Even if $\Gr\ll 1$, this can be superseded by $\frac{\rho_l}{\rho_v}\gg 1$ for larger droplets so that inertia dominates, hence the observed oscillatory relaxation. For smaller droplets, however, as $\Gr$ decreases drastically with $R$, it is the drag that come to dominate, hence a monotonic relaxation and the scenario of \S\ref{dropdyn}.

\section{Global picture Leidenfrost}
\label{global}

The larger the droplet is, the narrower the gap between the spherical droplet an the substrate becomes, as put into evidence by equation (\ref{eq:h}). As mentioned at the end of \S\ref{sec:limitations2}, a key limitation to the present analysis from the side of larger droplets is expected to be caused by a deviation from sphericity within such a narrow gap, even if the droplet as a whole might still remain largely spherical.   
As this region is crucial for vapour generation and heat transfer, in spite of its smallness,   
any morphological change therein can have a significant impact on the Leidenfrost phenomenon. 
On the other hand, it is through such a morphological change at the bottom of the droplet that a transition from the small spherical to larger non-spherical Leidenfrost droplets is bridged, which is touched upon in the present section. 

As established by~\cite{2012_Pomeau}, cf.~their equation (26), a significant deviation from sphericity at the bottom of the droplet occurs for $R$ starting from $R\sim\ell_i$, where the length scale $\ell_i$ is given by 
\begin{equation}
\ell_i=(\ell_*^3 \ell_c^4)^{1/7}
\label{elli}
\end{equation}
(all rewritten in our present notations). For $R\ll\ell_i$, the droplet is fully spherical and the analysis of the present paper holds. As $\ell_*\ll\ell_c$, equation (\ref{elli}) implies that $\ell_*\ll\ell_i\ll\ell_c$. For our reference case of a water droplet on a superheated substrate with $\Delta T=300^{\circ}$C \citep[][cf.~Appendix~\ref{app:Properties}]{2012_Celestini}, we obtain $\ell_i=367\,\mu\mathrm{m}$. 

\begin{figure}
  \centerline{\includegraphics[width=0.8\columnwidth]{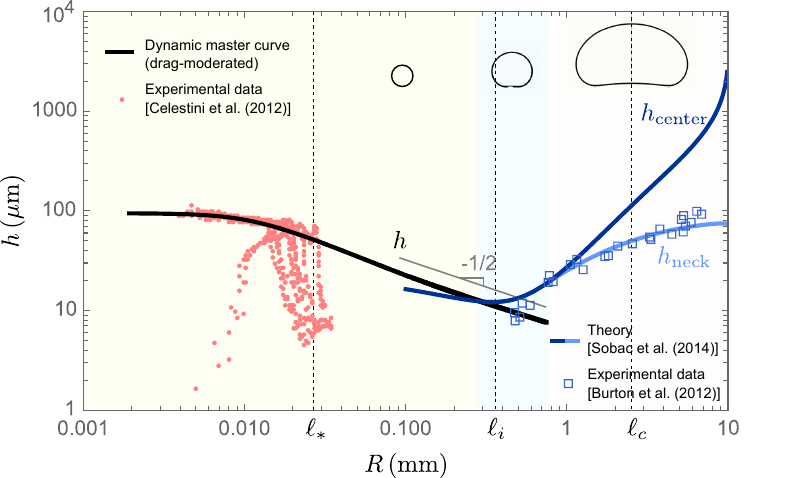}}
  \caption{Vapor film thickness $h$ under a water Leidenfrost droplet as a function of its radius $R$.  Theoretical predictions reproduced from~\cite{2014_Sobac,2021_Sobac_Erratum} for large deformed droplets ($R \gtrsim \ell_i$), and by the present model for small spherical droplets ($R\ll\ell_i$) are compared with experimental data from \cite{2012_Burton} and \cite{2012_Celestini}, respectively. Unlike the previous examples, the computations are here done with $\Delta T=270^{\circ}$C to follow \cite{2012_Burton} (cf.~also Appendix~\ref{app:Properties}), whereas the data by \cite{2012_Celestini} still correspond to $\Delta T=300^{\circ}$C (hence a slight misplacement relative to the theoretical curve as compared to previous figures). For large droplets, the $h$~curve splits into two branches $h_\text{neck}$ and $h_\text{center}$ at the point where the vapour layer between the droplet and the substrate adopts a `pocket-like' structure edged by a narrow annular neck, such that the minimum thickness no longer corresponds to the center and switches to the neck.}
\label{fig:Leidenfrost}
\end{figure}

Figure~\ref{fig:Leidenfrost} illustrates the Leidenfrost effect at large, over four decades of the droplet size. It combines the present results for the (small) spherical Leidenfrost droplets to the left of the figure (the dynamic master curve) with the results for the usual (large and deformed) droplets reproduced from \cite{2014_Sobac,2021_Sobac_Erratum} to the right. We note that for the non-spherical droplets the radius $R$ is here defined as the radius of the vertical projection on the substrate (i.e.~as the maximum horizontal radius). Two sets of experimental results, corresponding to the two drastically different size domains, are plotted alongside the theoretical curves: the ones by \cite{2012_Celestini} and by \cite{2012_Burton}. The overlapping occurs at $R\sim\ell_i$, quite as expected. Nevertheless, it does not appear to happen smoothly, which is most definitely due to some accuracy loss towards the limit of small near-spherical droplets within the model employed by \cite{2014_Sobac, 2021_Sobac_Erratum}. Notably, it is in this intermediate (overlapping) region that an absolute minimum of the vapour layer thicknesses is attained: from there, $h$ increases both towards the small spherical droplets as a take-off precursor and towards the large deformed droplets. 

\section{Conclusions}
The dynamics of small spherical Leidenfrost droplets have been investigated theoretically, yielding valuable new insights into their final stages of existence.

After numerically calculating the fluxes, evaporation rate, and forces for a spherical Leidenfrost droplet interacting with a superheated flat substrate (under the verifiable assumptions of quasi-stationarity and low Reynolds and P\'eclet numbers) and coming up with simple fitting formulas for these data as a function of the reduced height $\delta=h/R$ (all respecting the asymptotic behaviors as $\delta \to 0$ and $\delta\to\infty$, also studied here), a theoretical model has been developed which allows an accurate prediction of the droplet height $h$ as a function of the physical parameters without any fitting parameters.

First, by balancing the droplet's weight against the upward hydrodynamic force induced by evaporation, a `quasi-steady master curve' relating drop height $h$ to drop radius $R$ was derived. This  curve follow the $h\propto R^{-1/2}$ scaling law formulated by \cite{2012_Celestini,2012_Pomeau}. However, the prefactor is not constant  and rather varies from $3/2$ when $R\to\infty$ to $3/\sqrt{2}$ as $R \to 0$.

Furthermore, our analysis reveals that the aforementioned classical quasi-steady description, while capturing the general trend of the `take-off' phenomenon, is unable to accurately reproduce the experimental data of \cite{2012_Celestini}. Dynamic effects, especially those related to frictional forces, are crucial to accurately describe the take-off at small scales. Therefore, a `dynamic master curve', drag moderated, has been also derived and well reproduces the experimental data of \cite{2012_Celestini}. As a consequence of the friction effect, the Leidenfrost droplets disappear at a finite height, the value of which turns out to be universal for sufficiently large initial drops. This is in contrast to the prediction of the quasi-stationary model, which suggests an infinite final height. A formula that predicts this universal final height has been established.

Combining the present modeling (valid when $R<\ell_i$) with the one of \cite{2014_Sobac,2021_Sobac_Erratum} for larger deformed droplets (valid when $R>\ell_i$), we offer a comprehensive picture of the shape and elevation of Leidenfrost drops across a the full range of stable axisymmetric shapes, spanning four decades of drop sizes. These studies also align with the hierarchy of length scales $\ell_{*}<\ell_i<\ell_c$ pointed out by \cite{2012_Pomeau}, emphasizing the dominant physical mechanisms and associated scalings for each scale. 

In addition, a general dynamic model including both drag and inertia effects has been used to investigate the influence of initial conditions on droplet dynamics. For an initial height $h_0$ out of the equilibrium position $h_{\rm{QS}}$, a sufficiently large droplet approaches the `dynamic master curve' relatively quickly and in an oscillatory manner (like a damped oscillator), and then continues to evolve along this curve. Such small rebounds can indeed reproduce certain oscillatory trends observed in the experimental points of \cite{2012_Celestini}. This scenario holds for initial conditions not too far from equilibrium; otherwise, the droplets that find themselves too high at a given moment evaporate at a finite height before reaching the dynamic master curve, which highlights  a certain dependence of the Leidenfrost droplet deposition manner on the result. 

We hope this research will stimulate further theoretical investigations, particularly in the intermediate region ($R\approx \ell_i$) where the vapor layer is thinnest, and encourage further experimental studies on Leidenfrost droplets with $R\lesssim\ell_i$, an area that remains underexplored.

\subsection*{Acknowledgments}
BS gratefully acknowledges the support from Centre National de la Recherche Scientifique -- CNRS, AR from BELSPO and ESA PRODEX Evaporation, and PC from the Fonds de la Recherche Scientifique -- FNRS.

\appendix

\section{Properties}
\label{app:Properties}

The properties used in this work are provided in table~\ref{tab:properties} \citep{2015_Sobac,2017_Sobac}.
\begin{table*}
\begin{tabular}{l c c c c c c c c c c c c c}
\hline
  & $T_{\rm sat}\, $ &  $\rho_l$ & $\L$  & $\gamma$  & $\ell_c$ & $\Delta T$ & $\lambda_v$  &  $\mu_v$ & $\rho_v$ & $\ell_*$ & $\ell_i$ & $\epsilon$ & $\mathcal{E}\times 10^6$ \\ 
Liquid &  ($^\circ$C)  &  $\left(\mathrm{\frac{kg}{m^3}}\right)$ & $\left(\mathrm{\frac{kJ}{kg}}\right)$  & $\left(\mathrm{\frac{mN}{m}}\right)$   & (mm) & ($^\circ$C) & $\left(\mathrm{\frac{m\!W}{m\,K}}\right)$ & ($\mu$Pa$\,$s) & $\left(\mathrm{\frac{kg}{m^3}}\right)$ &  ($\mu$m) & ($\mu$m)  & (--) & (--) \\
\hline
Water  & 100 & 960 & 2555 & 59 & 2.50 & 300 & 36.9 & 18.6  & 0.42 & 28.5 & 367  & 0.076 &  1.47 \\ 
Water  & 100 & 960 & 2555 & 59 & 2.50 & 270 & 35.8 & 18.0  & 0.43 & 26.7  & 357 & 0.077 & 1.21   \\
Ethanol & 79 & 728 & 840 & 17 & 1.56 & 321   & 25.4 & 15.6 & 1.10  & 26.9  & 274  & 0.115  & 5.12  \\
Ethanol & 79 & 728 & 840 & 17 & 1.56 & 217   & 22.7 & 14.0 & 1.22 &  21.1 & 247  &  0.119  &  2.48  \\
\hline 
\end{tabular}
\caption{Parameter values at $1\,\mathrm{atm}$ (where relevant). Liquid density $\rho_l$, latent heat $\L$, surface tension $\gamma$ and capillary length $\ell_c=\sqrt{\gamma/\rho_l g}$ ($g$ gravity acceleration) at the boiling temperature $T_{\rm sat}$; superheat $\Delta T\equiv T_w-T_\text{sat}$, with $T_w$ the substrate temperatrue; vapor thermal conductivity $\lambda_v$, viscosity $\mu_v$, density $\rho_v$ at the mid-temperature $\frac{1}{2}(T_{\rm sat}+ T_w)$; take-off scale $\ell_*$, cf.~(\ref{scales1}); non-sphericity scale $\ell_i$, cf.~(\ref{elli}); $\epsilon\equiv (\rho_v/\rho_l)^{1/3}$; evaporation number $\mathcal{E}$, cf.~(\ref{evapnumber}). The first and third cases (raws) correspond to \cite{2012_Celestini}, while the second one to \cite{2012_Burton} and the fourth one to \cite{2019_Lyu}.}
\label{tab:properties}
\end{table*} 

To facilitate continuity with some previous studies of the usual (larger non-spherical) Leidenfrost droplets~\citep{2014_Sobac,2021_Sobac_Erratum,2021_Chantelot}, apart from the parameters defined in the main text, we here also follow the (dimensionless) evaporation number 
\begin{equation}
\E= \frac{\mu_v \lambda_v \Delta T}{\gamma \L \rho_v \ell_c} \,.
\label{evapnumber}
\end{equation}
We note the formulas $\ell_*=\E^{1/3} \ell_c$ and $\ell_i=\E^{1/7} \ell_c$ resulting from (\ref{scales1}), (\ref{elli}) and (\ref{evapnumber}).

\section{Some more precise fits of numerical data}
\label{app:fits}

In addition to the simplified fits outlined in the main text, more precise fits are proposed herein. These fits are presented alongside and compared with the numerically computed data in Fig.\ref{fig:Fit_Asymp_Comp}. The asymptotic behaviors derived in Appendix~\ref{app:Asymptotic} are also depicted.

 \begin{figure}
  \centerline{\includegraphics[width=\columnwidth]{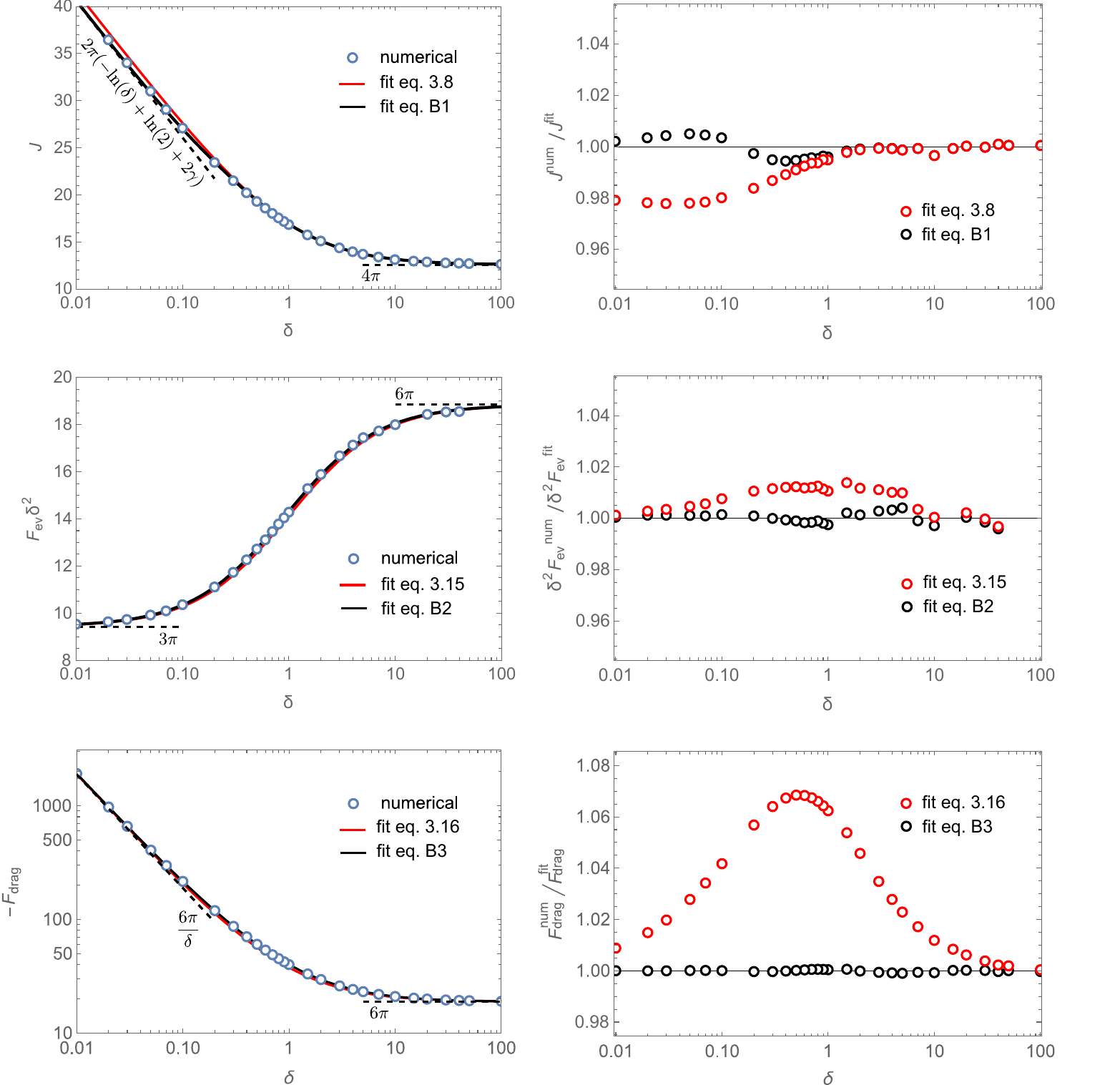}}
  \caption{(Left) Evaporation rate of and forces acting on a spherical Leidenfrost droplet as a function of its reduced height $\delta$. Numerically computed data are presented alongside  their asymptotic behaviours and compared with fits of two different levels of precision. (Right) Respective ratio of the numerical data to the proposed fits as a function of $\delta$.}
\label{fig:Fit_Asymp_Comp}
\end{figure}

\subsection{Fit for $J(\delta)$}
\label{app:Jfit}
The fit (\ref{eq:FitJbis}) can further be improved as follows: 
\begin{equation}
 J(\delta)= 4\pi\left[1+\ln\left(1+\frac{1}{\delta}\right)- \left(1-\frac{1}{2}\ln 2-\gamma\right) \frac{1}{1+50.8 \delta^2} \right]\,,
   \label{eq:FitJ}
\end{equation}
where $\gamma=0.577216$ is the Euler constant. The expression (\ref{eq:FitJ}) respects the two-term asymptotic expansions both as $\delta\to 0$ and as $\delta\to +\infty$ obtained in Appendix~\ref{Jasy0} and Appendix~\ref{Jasyinf}, respectively.  

\subsection{Fit for $F_\text{ev}(\delta)$}
\label{app:Fev}
A more precise fit than (\ref{eq:FitFevbis}) is given by 
\begin{equation}
 F_{\rm{ev}}(\delta)= \frac{3 \pi}{\delta^2} \left(1 + \frac{\delta}{0.924 + \delta} \right)\,.
  \label{eq:FitFev}
\end{equation}

\subsection{Fit for $F_\text{drag}(\delta)$}
\label{app:DragForce}

A better fit than (\ref{eq:FitFdragbis}) is provided by the formula 
\begin{equation}
F_\text{drag}(\delta)=6\pi\left(1+ \frac{1}{\delta}+1.161 \frac{1+26.01\delta}{1+62.447\delta+187.12\delta^2+2.514 \delta^3} \right)\,. 
\label{eq:FitFdrag}
\end{equation} 

\section{Asymptotic behaviours}
\label{app:Asymptotic}

\subsection{$J$ as $\delta\to 0$}
\label{Jasy0}
When the sphere (dimensionless radius unity) is close to the substrate ($\delta\ll 1$), its profile $z=h(r)$ ($z=0$ at the substrate) in a small vicinity of the downmost point is well approximated by a parabola 
$h=\delta +\frac{1}{2} r^2$ matching with the outer circular drop shape. 
The evaporation flux is given by heat conduction across this thin vapour layer, viz.\ $j=1/h$ in dimensionless form. Integrating over such a small vicinity up to a reference point $r=r_1\ll 1$, one obtains 
$J_1=2\pi\int_0^{r_1}r\, j\,\mathtt{d}r =2\pi \ln(\delta+\frac{1}{2} r_1^2)-2\pi \ln\delta$ 
for the first contribution into the evaporation rate $J$. One can observe a logarithmic divergence and assume the sought asymptotic behaviour in the form $J\sim 2\pi (-\ln\delta+\mathrm{const})$ as $\delta\to 0$. However, to fully determine $\mathrm{const}$, one needs to consider the contribution $J_2$ from the rest of the sphere, such that eventually $J=J_1+J_2$. To this purpose, it suffices to solve the problem (\ref{eq:lapl})--(\ref{eq:BCT3}) for a unit sphere lying on the substrate (with formally $\delta=0$). This can be done numerically with $J_2$ determined by integrating (\ref{eq:evaporation_flux}) over the rest of the sphere up to the point $r=r_1$ on the lower surface. The result diverges in the limit $r_1\to 0$: as $J_2=-4\pi\ln r_1+\mathrm{const}_1$, where $\mathrm{const}_1$ is determined numerically by choosing sufficiently small $r_1$. In this way, we finally arrive at $J\sim 2\pi (-\ln\delta+1.85)$ as $\delta\to 0$ (the terms with $r_1$ cancelling out in $J_1+J_2$). Within the computation precision and with $\gamma$ being the Euler constant, this can be rewritten as $J\sim 2\pi (-\ln\delta+\ln 2+2\gamma)$ as $\delta\to 0$, which is taken into account when constructing the fit (\ref{eq:FitJ}). This (exact) value of the constant can in principle be obtained from the exact solution of the present heat conduction problem in curvilinear coordinates~~\citep[similarly to e.g.~][]{Lebedev, Popov2005}, although we here limit ourselves to corresponding numerical solutions. The simplified fit (\ref{eq:FitJbis}) respects exactly the logarithmic divergence,  but just approximately the constant. 

\subsection{$J$ as $\delta\to +\infty$}
\label{Jasyinf}
When the sphere is far away from the substrate, the leading-order result is as for the sphere in an unbounded medium: 
\begin{equation}
T=1-1/\text{r}\,,
\label{Torig}
\end{equation}
where $\text{r}=\sqrt{r^2+(z-\delta-1)^2}$ is the spherical radial coordinate from the centre of the sphere (note the font difference with the cylindrical radial coordinate $r$). Therefore, $j=\partial_{\text{r}}T|_{\text{r}=1}=1$ and hence $J=4\pi$. 
The first correction comes from the sphere reflection in the substrate. The image sphere adds the following primary contribution into the temperature field in the original domain (above the substrate, $z>0$):  
\begin{equation}
T_\text{im}=1/\text{r}_\text{im}\,,
\label{Tim}
\end{equation}
where $\text{r}_\text{im}=\sqrt{r^2+(z+\delta+1)^2}$ and the subscript `im' is associated with the image.  
The primary effect of (\ref{Tim}) is to increase the local ambient temperature around the original sphere from $T=1$ to $T=1+1/(2\delta)$. As $T=0$ at the sphere surface, cf.~(\ref{eq:BCT3}), the values of $j$ and $J$ are then increased in the same proportion. Thus, we arrive at $J\sim 4\pi (1+1/(2\delta))$ as $\delta\to +\infty$, which is respected by both the fit (\ref{eq:FitJbis}) and (\ref{eq:FitJ}). 

\subsection{$F_{\rm{ev}}$ as $\delta\to 0$}
\label{Fevasy0}
In dimensionless terms (scales provided in table~\ref{tab:scales}), the lubrication equation in the thin vapour layer between the substrate and the sphere can be written as $\frac{1}{12 r}\partial_r (r h^3 \partial_r P_v)+\frac{1}{h}=0$ \citep[cf.~also][]{2021_Sobac_Erratum}, where $P_v$ is the vapour pressure excess over the ambient one (hence $P_v\to 0$ far away). Integrating with $h=\delta +\frac{1}{2} r^2$ and on account of symmetry $\partial_r P_v|_{r=0}=0$, one obtains $\partial_r P_v=12 h^{-3} \ln\frac{h}{\delta}$. The leading-order contribution into $F_\text{ev}$ is given by $F_\text{ev}=2\pi\int_0^{+\infty} r P_v \mathtt{d}r=\pi r^2 P_v|_{r\to+\infty}-\pi \int_0^{+\infty} r^2 \partial_r P_v \mathtt{d}r=3\pi/\delta^2$, which is the sought behaviour as $\delta\to 0$ and is respected in (\ref{eq:FitFev}). 

\subsection{$F_{\rm{ev}}$ as $\delta\to +\infty$}
\label{Fevasyinf}
Assuming $\delta\gg 1$, we shall distinguish three contributions into the sough asymptotic behaviour: $F_\text{ev}=F_\text{ev1}+F_\text{ev2}+F_\text{ev3}$, which are all of the same order. 

First, the leading-order (spherically symmetric) flow field 
\begin{equation}
\vv=\frac{\textbf{r}}{\text{r}^3}
\label{vorig}
\end{equation}
from our evaporating  sphere ($\textbf{r}$ being the position vector from the sphere centre) is supplemented by the one from the image sphere 
\begin{equation}
\vv_\text{im}=\frac{\textbf{r}_\text{im}}{\text{r}_\text{im}^3}
\label{vim}
\end{equation}
(in the original domain $z>0$). At the location of the original sphere ($r=0$, $z=\delta+1$), in the limit $\delta\gg 1,$ the velocity field (\ref{vim}) gives rise to a (quasi-)uniform streaming velocity $v_0=1/(4(\delta+1)^2)\sim 1/(4\delta^2)$ directed vertically upwards. This in turn gives rise to the Stokes drag $6\pi \mu_v R\, v_0$ upon the original sphere (in dimensional terms). In our present dimensionless terms (cf.~table~\ref{tab:scales}), this amounts to $F_\text{ev1}=6\pi v_0=3\pi/(2\delta^2)$.  

Second, the superposition of the velocity fields (\ref{vorig}) and (\ref{vim}) does satisfy the impermeability condition at the substrate: $v_z=0$ at $z=0$ (hereafter $v_r$ and $v_z$ are the $r$- and $z$-components of the velocity field). However, the no-slip condition $v_r=0$ at $z=0$ is not satisfied. To remedy this, we consider another contribution into the velocity field, the addition of which permits to observe the no-slip condition. We proceed in terms of the stream function $\psi$: 
\begin{equation}
v_r=\frac{1}{r} \partial_z \psi\,,\quad v_z=-\frac{1}{r} \partial_r \psi\,.
\label{psi}
\end{equation}
The Stokes equation can be written as \citep[cf.~e.g.][]{HappelBrenner} 
\begin{equation}
E^2 E^2\psi=0\,, \quad E^2= \partial_{rr}-\frac{1}{r} \partial_r + \partial_{zz} \,,
\label{Stokeseq}
\end{equation}
which is solved in the domain $z>0$ with the boundary conditions  
\begin{equation}
\psi=0\,,\ \partial_z\psi=-\frac{2 r^2}{[(\delta+1)^2+r^2]^{3/2}}\ \text{at}\ z=0 \,, \quad
\psi/\text{r}^2\to 0 \ \text{at\ infinity}\,.
\label{BCs}
\end{equation}
The slip velocity in the second condition (\ref{BCs}) is directed towards the axis and is such as to offset the corresponding contribution from the sum of (\ref{vorig}) and (\ref{vim}). One can verify that 
\begin{equation}
\psi=-\frac{2 r^2 z}{[(\delta+1+z)^2+r^2]^{3/2}}
\label{sol}
\end{equation}
is an exact solution of the problem (\ref{Stokeseq}), (\ref{BCs}). Eventually, we are just interested in the velocity field value $v_0$ at the location of the original sphere ($r=0$, $z=\delta+1$). Using (\ref{psi}) and (\ref{sol}), one obtains $v_0=1/(2(\delta+1)^2)\sim 1/(2\delta^2)$ directed vertically upwards. As with the first contribution, the Stokes drag considerations lead to the result $F_\text{ev2}=6\pi v_0=3\pi/\delta^2$. 

Third, the temperature field (\ref{Tim}) from the image sphere gives rise not only to an effective uniform temperature increase in the original sphere surrounding, already taken into account in Appendix~\ref{Jasyinf}, but also to a (dimensionless) temperature gradient $\partial_z T \sim -1/(4\delta^2)$. This breaks down the spherical symmetry of the evaporation flux ($j$ no longer constant along the sphere surface) and of the evaporative flow (a correction upon (\ref{vorig})). Hydrodynamically, this can engender an additional force contribution $F_\text{ev3}$. We proceed with the analysis using the spherical coordinates $\{\text{r},\theta\}$ related to the original sphere, such that $r=\text{r} \sin\theta$ and $z=\delta+1+\text{r} \cos\theta$. Then the problem for the mentioned (gradient-related) part of the temperature field around the original sphere can be formulated as 
\begin{eqnarray}
\nabla^2 T=0\,,\quad \nabla^2=\partial_{\text{r}\text{r}}+\frac{2}{\text{r}} \partial_\text{r} + \frac{1}{\text{r}^2 \sin\theta} \partial_\theta \sin\theta \,\partial_\theta\,,
\label{Teq2}\label{lapl2}\\
T=0 \ \ \text{at}\ \text{r}=1\,,\quad T\sim -\frac{1}{4\delta^2} \text{r} \cos\theta \ \ \text{as}\ \text{r}\to +\infty\, \label{BCs2}
\end{eqnarray}
to be solved in the domain $\text{r}>1$. The infinity in (\ref{BCs2}) formally corresponds to $1\ll \text{r}\ll\delta$. 
The solution of (\ref{lapl2}) with (\ref{BCs2}) is 
\begin{equation}
T=-\frac{1}{4\delta^2} \left(\text{r} - \frac{1}{\text{r}^2}\right) \cos\theta\,. 
\label{sol2}
\end{equation}
Therefore, 
\begin{equation}
j=\partial_{\text{r}}T|_{\text{r}=1}=-\frac{3}{4\delta^2} \cos\theta\,, 
\label{j2}
\end{equation}
which shows that the present contribution corresponds to evaporation reduction at the upper part of the sphere ($j<0$ for $0\le\theta<\pi/2$) and intensification at the lower part of the sphere ($j>0$ for $\pi/2<\theta\le \pi$), closer to the substrate, as expected. 
To calculate the flow induced by (\ref{j2}), we work once again in terms of the stream function, now in the spherical coordinates:   
\begin{equation}
v_\text{r}=\frac{1}{\text{r}^2\sin\theta} \partial_\theta \psi\,,\quad v_\theta=-\frac{1}{\text{r}\sin\theta} \partial_\text{r} \psi\,
\label{psi2}
\end{equation}
for the $\text{r}$- and $\theta$- components of the velocity field. The problem is formulated in the domain $\text{r}>1$. The Stokes equation \citep[cf.~e.g.][]{HappelBrenner} and the boundary conditions can be written as 
\begin{eqnarray}
E^2 E^2 \psi=0\,,\quad E^2=\partial_{\text{r}\text{r}} + \frac{\sin\theta}{\text{r}^2} \partial_\theta \frac{1}{\sin\theta} \,\partial_\theta\,,
\label{Teq2}\label{Stokeseq2}\\
\psi=-\frac{3}{8\delta^2} \sin^2\theta\,,\ \partial_\text{r}\psi=0 \ \ \text{at}\ \text{r}=1\,,\quad \psi/\text{r}^2\to 0 \ \ \text{as}\ \text{r}\to +\infty\,. \label{BCs2bis}
\end{eqnarray}
The first condition (\ref{BCs2bis}) corresponds to $v_r=j$ (in our dimensionless terms) on account of (\ref{j2}) and (\ref{psi2}), while the second condition (\ref{BCs2bis}) to no slip ($v_\theta=0$). The solution of (\ref{Stokeseq2}) with (\ref{BCs2bis}) is 
\begin{equation}
\psi=-\frac{3}{8\delta^2} \left(\text{r}+\frac{1}{\text{r}}\right) \frac{\sin^2\theta}{2}\,.
\label{sol2bis}
\end{equation}
The force acting on the sphere in the $z$-direction is determined by $(-4\pi)$ times the `Stokeslet' prefactor, i.e. the one at the term $\text{r}\frac{\sin^2\theta}{2}$ \citep[cf.][]{HappelBrenner}. Thus, from (\ref{sol2bis}), one obtains $F_\text{ev3}=3\pi/(2\delta^2)$. 

Summing up the three contributions, one finally obtains $F_\text{ev}=6\pi/\delta^2$ (as $\delta\to +\infty$). It is noteworthy that the asymptotic behaviour remains $O(\delta^{-2})$ in both limits: as $\delta\to +\infty$ and as $\delta\to 0$ (cf.~Appendix~\ref{Fevasy0}). Yet, the prefactors are twice different. 

\section{Case of ethanol}
\label{app:Ethanol}

\begin{figure}
  \centerline{\includegraphics[width=\columnwidth]{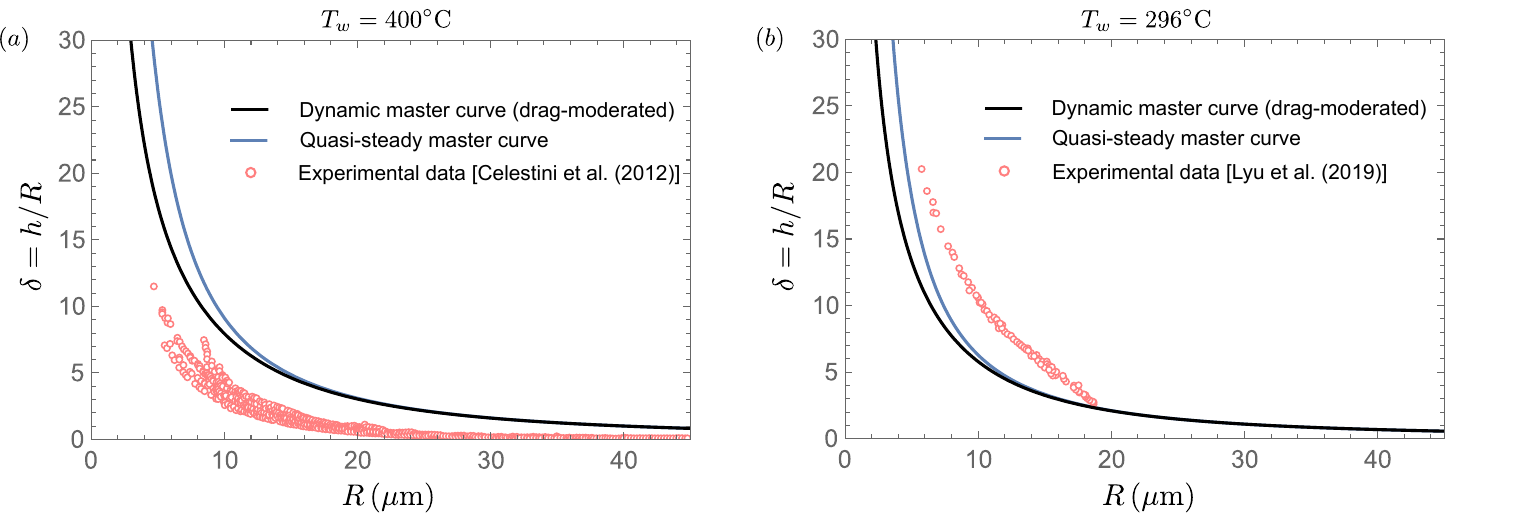}}
  \caption{Relative height $\delta=h/R$ of ethanol droplets as a function of $R$. The dynamic and quasi-steady master curves are here compared to the experimental data by \cite{2012_Celestini} to the left and \cite{2019_Lyu} to the right, cf.~table~\ref{tab:properties} for the parameter values.}
\label{fig:Ethanol}
\end{figure}

Since \cite{2012_Celestini} and \cite{2019_Lyu} have conducted experiments with ethanol droplets, it is worthwhile to compare the present theoretical predictions with those experimental results. Although the predictions match the form and order of magnitude of the experimental results, the agreement is less satisfactory than in the case of water. Specifically, we overestimate the data from \cite{2012_Celestini} and underestimate the data from \cite{2019_Lyu}, cf.~figure~\ref{fig:Ethanol}. Such multidirectional discrepancies make it challenging to propose a hypothesis that could explain the differences, which might be attributed to experimental errors or some physical ingredient missing in the model that is important for ethanol, but not for water. Further investigation (including experimental) is needed to address this issue. 

\newpage

\bibliography{Leidenfrost}

\begin{thebibliography}{40}
\expandafter\ifx\csname natexlab\endcsname\relax\def\natexlab#1{#1}\fi
\def\au#1{#1} \def\ed#1{#1} \def\yr#1{#1}\def\at#1{#1}\def\jt#1{\textit{#1}}
  \def\bt#1{#1}\def\bvol#1{\textbf{#1}} \def\vol#1{#1} \def\pg#1{#1}
  \def\publ#1{#1}\def\arxiv#1{#1}\def\org#1{#1}\def\st#1{\textit{#1}}

\bibitem[Ajaev \& Kabov(2021)]{2021_Ajaev}
{\sc \au{Ajaev, Vladimir~S} \& \au{Kabov, Oleg~A}} \yr{2021}  \at{Levitation
  and self-organization of droplets}.  \jt{Annual Review of Fluid Mechanics}
  \bvol{53},  \pg{203--225}.

\bibitem[Baier {\em et~al.\/}(2013)Baier, Dupeux, Herbert, Hardt \&
  Qu{\'e}r{\'e}]{2013_Baier}
{\sc \au{Baier, Tobias}, \au{Dupeux, Guillaume}, \au{Herbert, Stefan},
  \au{Hardt, Steffen} \& \au{Qu{\'e}r{\'e}, David}} \yr{2013}  \at{Propulsion
  mechanisms for leidenfrost solids on ratchets}.  \jt{Physical Review E}
  \bvol{87}~(2),  \pg{021001}.

\bibitem[Bergen {\em et~al.\/}(2019)Bergen, Basso \& Bostwick]{2019_Bergen}
{\sc \au{Bergen, Jesse~E}, \au{Basso, Bailey~C} \& \au{Bostwick, Joshua~B}}
  \yr{2019}  \at{Leidenfrost drop dynamics: Exciting dormant modes}.
  \jt{Physical Review Fluids}  \bvol{4}~(8),  \pg{083603}.

\bibitem[Biance {\em et~al.\/}(2003)Biance, Clanet \&
  Qu{\'e}r{\'e}]{2003_Biance}
{\sc \au{Biance, Anne-Laure}, \au{Clanet, Christophe} \& \au{Qu{\'e}r{\'e},
  David}} \yr{2003}  \at{Leidenfrost drops}.  \jt{Phys. Fluids}  \bvol{15}~(6),
   \pg{1632--1637}.

\bibitem[Boerhaave(1732)]{Boerhaave1732}
{\sc \au{Boerhaave, H.}} \yr{1732} {\em Elementae Chemiae, Vol. 1.\/}.
  \publ{Leiden: Lugduni Batavorum}.

\bibitem[Bouillant {\em et~al.\/}(2021{\natexlab{{\em a\/}}})Bouillant, Cohen,
  Clanet \& Qu{\'e}r{\'e}]{2021_Bouillant_PNAS}
{\sc \au{Bouillant, Ambre}, \au{Cohen, Caroline}, \au{Clanet, Christophe} \&
  \au{Qu{\'e}r{\'e}, David}} \yr{2021{\natexlab{{\em a\/}}}}
  \at{Self-excitation of leidenfrost drops and consequences on their
  stability}.  \jt{Proceedings of the National Academy of Sciences}
  \bvol{118}~(26),  \pg{e2021691118}.

\bibitem[Bouillant {\em et~al.\/}(2021{\natexlab{{\em b\/}}})Bouillant, Lafoux,
  Clanet \& Qu{\'e}r{\'e}]{2021_Bouillant_SM}
{\sc \au{Bouillant, Ambre}, \au{Lafoux, Baptiste}, \au{Clanet, Christophe} \&
  \au{Qu{\'e}r{\'e}, David}} \yr{2021{\natexlab{{\em b\/}}}}  \at{Thermophobic
  leidenfrost}.  \jt{Soft Matter}  \bvol{17}~(39),  \pg{8805--8809}.

\bibitem[Bouillant {\em et~al.\/}(2018)Bouillant, Mouterde, Bourrianne,
  Lagarde, Clanet \& Qu{\'e}r{\'e}]{2018_Bouillant}
{\sc \au{Bouillant, Ambre}, \au{Mouterde, Timoth{\'e}e}, \au{Bourrianne,
  Philippe}, \au{Lagarde, Antoine}, \au{Clanet, Christophe} \&
  \au{Qu{\'e}r{\'e}, David}} \yr{2018}  \at{Leidenfrost wheels}.  \jt{Nature
  Physics}  \bvol{14}~(12),  \pg{1188--1192}.

\bibitem[Brunet \& Snoeijer(2011)]{2011_Brunet}
{\sc \au{Brunet, P.} \& \au{Snoeijer, J.~H.}} \yr{2011}  \at{Star-drops formed
  by periodic excitation and on air cushion - a short review}.  \jt{Eur. Phys.
  J. Special Topics}  \bvol{192},  \pg{207--226}.

\bibitem[Brutin(2015)]{2015_Brutin}
{\sc \au{Brutin, David}}, ed. \yr{2015} {\em Droplet wetting and evaporation:
  from pure to complex fluids\/}.  \publ{Oxford: Academic Press}.

\bibitem[Burton {\em et~al.\/}(2012)Burton, Sharpe, van~der Veen, Franco \&
  Nagel]{2012_Burton}
{\sc \au{Burton, J.~C.}, \au{Sharpe, A.~L.}, \au{van~der Veen, R. C.~A.},
  \au{Franco, A.} \& \au{Nagel, S.~R.}} \yr{2012}  \at{Geometry of the vapor
  layer under a leidenfrost drop}.  \jt{Phys. Rev. Lett.}  \bvol{109},
  \pg{074301}.

\bibitem[Celestini {\em et~al.\/}(2012)Celestini, Frisch \&
  Pomeau]{2012_Celestini}
{\sc \au{Celestini, Franck}, \au{Frisch, Thomas} \& \au{Pomeau, Yves}}
  \yr{2012}  \at{Take off of small leidenfrost droplets}.  \jt{Phys. Rev.
  Lett.}  \bvol{109},  \pg{034501}.

\bibitem[Chakraborty {\em et~al.\/}(2022)Chakraborty, Chubynsky \&
  Sprittles]{2022_Chakraborty}
{\sc \au{Chakraborty, Indrajit}, \au{Chubynsky, Mykyta~V} \& \au{Sprittles,
  James~E}} \yr{2022}  \at{Computational modelling of leidenfrost drops}.
  \jt{Journal of Fluid Mechanics}  \bvol{936},  \pg{A12}.

\bibitem[Chantelot \& Lohse(2021)]{2021_Chantelot}
{\sc \au{Chantelot, Pierre} \& \au{Lohse, Detlef}} \yr{2021}  \at{Drop impact
  on superheated surfaces: short-time dynamics and transition to contact}.
  \jt{Journal of Fluid Mechanics}  \bvol{928},  \pg{A36}.

\bibitem[Dodd {\em et~al.\/}(2020)Dodd, Agrawal, Geraldi, Xu, Wells, Martin,
  Newton, McHale \& Wood]{2020_Dodd}
{\sc \au{Dodd, Linzi~E}, \au{Agrawal, Prashant}, \au{Geraldi, Nicasio~R},
  \au{Xu, Ben~B}, \au{Wells, Gary~G}, \au{Martin, James}, \au{Newton,
  Michael~I}, \au{McHale, Glen} \& \au{Wood, David}} \yr{2020}  \at{Planar
  selective leidenfrost propulsion without physically structured substrates or
  walls}.  \jt{Applied Physics Letters}  \bvol{117}~(8).

\bibitem[Dodd {\em et~al.\/}(2019)Dodd, Agrawal, Parnell, Geraldi, Xu, Wells,
  Stuart-Cole, Newton, McHale \& Wood]{2019_Dodd}
{\sc \au{Dodd, Linzi~E}, \au{Agrawal, Prashant}, \au{Parnell, Matthew~T},
  \au{Geraldi, Nicasio~R}, \au{Xu, Ben~B}, \au{Wells, Gary~G}, \au{Stuart-Cole,
  Simone}, \au{Newton, Michael~I}, \au{McHale, Glen} \& \au{Wood, David}}
  \yr{2019}  \at{Low-friction self-centering droplet propulsion and transport
  using a leidenfrost herringbone-ratchet structure}.  \jt{Physical Review
  Applied}  \bvol{11}~(3),  \pg{034063}.

\bibitem[Dupeux {\em et~al.\/}(2011)Dupeux, Le~Merrer, Clanet \&
  Qu{\'e}r{\'e}]{2011_Dupeux}
{\sc \au{Dupeux, Guillaume}, \au{Le~Merrer, Marie}, \au{Clanet, Christophe} \&
  \au{Qu{\'e}r{\'e}, David}} \yr{2011}  \at{Trapping leidenfrost drops with
  crenelations}.  \jt{Phys. Rev. Lett.}  \bvol{107},  \pg{114503}.

\bibitem[Graeber {\em et~al.\/}(2021)Graeber, Regulagadda, Hodel, K{\"u}ttel,
  Landolf, Schutzius \& Poulikakos]{2021_Graeber}
{\sc \au{Graeber, Gustav}, \au{Regulagadda, Kartik}, \au{Hodel, Pascal},
  \au{K{\"u}ttel, Christian}, \au{Landolf, Dominic}, \au{Schutzius, Thomas~M}
  \& \au{Poulikakos, Dimos}} \yr{2021}  \at{Leidenfrost droplet trampolining}.
  \jt{Nature communications}  \bvol{12}~(1),  \pg{1727}.

\bibitem[Guyon {\em et~al.\/}(2012)Guyon, Hulin \&
  Petit]{guyon2012hydrodynamique}
{\sc \au{Guyon, Etienne}, \au{Hulin, Jean-Pierre} \& \au{Petit, Luc}} \yr{2012}
  {\em Hydrodynamique physique 3e {\'e}dition\/}.  \publ{EDP sciences}.

\bibitem[Happel \& Brenner(1965)]{HappelBrenner}
{\sc \au{Happel, J.} \& \au{Brenner, H.}} \yr{1965} {\em Slow Reynolds Number
  Hydrodynamics\/}.  \publ{Prentice-Hall}.

\bibitem[Lebedev(1972)]{Lebedev}
{\sc \au{Lebedev, N.N.}} \yr{1972} {\em Special Functions and Their
  Applications\/}.  \publ{Revised edition, Dover Publications}.

\bibitem[Leidenfrost(1756)]{1756Leidenfrost}
{\sc \au{Leidenfrost, J.~G.}} \yr{1756} {\em De Aquae Communis Nonnullis
  Qualitibus Tractatus, Part 2\/}.  \publ{Duisburg}.

\bibitem[Li {\em et~al.\/}(2023)Li, Li, Lyu, Zhao, Xue, Li, Li, Li, Sun \&
  Song]{2023_Li}
{\sc \au{Li, An}, \au{Li, Huizeng}, \au{Lyu, Sijia}, \au{Zhao, Zhipeng},
  \au{Xue, Luanluan}, \au{Li, Zheng}, \au{Li, Kaixuan}, \au{Li, Mingzhu},
  \au{Sun, Chao} \& \au{Song, Yanlin}} \yr{2023}  \at{Tailoring vapor film
  beneath a leidenfrost drop}.  \jt{Nature Communications}  \bvol{14}~(1),
  \pg{2646}.

\bibitem[Linke {\em et~al.\/}(2006)Linke, Alem{\'a}n, Melling, Taormina,
  Francis, Dow-Hygelund, Narayanan, Taylor \& Stout]{2006_Linke}
{\sc \au{Linke, H.}, \au{Alem{\'a}n, B.~J.}, \au{Melling, L.~D.}, \au{Taormina,
  M.~J.}, \au{Francis, M.~J.}, \au{Dow-Hygelund, C.~C.}, \au{Narayanan, V.},
  \au{Taylor, R.~P.} \& \au{Stout, A.}} \yr{2006}  \at{Self-propelled
  leidenfrost droplets}.  \jt{Phys. Rev. Lett.}  \bvol{96}~(154502).

\bibitem[Liu \& Tran(2020)]{2020_Liu}
{\sc \au{Liu, Dongdong} \& \au{Tran, Tuan}} \yr{2020}  \at{Size-dependent
  spontaneous oscillations of leidenfrost droplets}.  \jt{Journal of Fluid
  Mechanics}  \bvol{902},  \pg{A21}.

\bibitem[Lyu {\em et~al.\/}(2019)Lyu, Mathai, Wang, Sobac, Colinet, Lohse \&
  Sun]{2019_Lyu}
{\sc \au{Lyu, Sijia}, \au{Mathai, Varghese}, \au{Wang, Yujie}, \au{Sobac,
  Benjamin}, \au{Colinet, Pierre}, \au{Lohse, Detlef} \& \au{Sun, Chao}}
  \yr{2019}  \at{Final fate of a leidenfrost droplet: Explosion or takeoff}.
  \jt{Science advances}  \bvol{5}~(5),  \pg{eaav8081}.

\bibitem[Ma \& Burton(2018)]{2018_Ma}
{\sc \au{Ma, Xiaolei} \& \au{Burton, Justin~C}} \yr{2018}  \at{Self-organized
  oscillations of leidenfrost drops}.  \jt{Journal of Fluid Mechanics}
  \bvol{846},  \pg{263--291}.

\bibitem[Ma {\em et~al.\/}(2017)Ma, Li{\'e}tor-Santos \& Burton]{2017_Ma}
{\sc \au{Ma, Xiaolei}, \au{Li{\'e}tor-Santos, Juan-Jos{\'e}} \& \au{Burton,
  Justin~C}} \yr{2017}  \at{Star-shaped oscillations of leidenfrost drops}.
  \jt{Physical Review Fluids}  \bvol{2}~(3),  \pg{031602}.

\bibitem[Marengo \& De~Coninck(2022)]{2022_Marengo}
{\sc \au{Marengo, Marco} \& \au{De~Coninck, Joel}}, ed. \yr{2022} {\em The
  Surface Wettability Effect on Phase Change\/}.  \publ{Cham: Springer
  International Publishing}.

\bibitem[Mar{\'\i}n {\em et~al.\/}(2012)Mar{\'\i}n, Arnaldo~del Cerro,
  R{\"o}mer, Pathiraj, Lohse {\em et~al.\/}]{2012_Marin}
{\sc \au{Mar{\'\i}n, {\'A}lvaro~G}, \au{Arnaldo~del Cerro, Daniel},
  \au{R{\"o}mer, Gertwillem~RBE}, \au{Pathiraj, B}, \au{Lohse, Detlef} \&
  \au{others}} \yr{2012}  \at{Capillary droplets on leidenfrost
  micro-ratchets}.  \jt{Physics of fluids}  \bvol{24}~(12).

\bibitem[Pomeau {\em et~al.\/}(2012)Pomeau, Le~Berre, Celestini \&
  Frisch]{2012_Pomeau}
{\sc \au{Pomeau, Yves}, \au{Le~Berre, Martine}, \au{Celestini, Franck} \&
  \au{Frisch, Thomas}} \yr{2012}  \at{The leidenfrost effect: From
  quasi-spherical droplets to puddles}.  \jt{Comptes Rendus Mecanique}
  \bvol{340}~(11-12),  \pg{867--881}.

\bibitem[Popov(2005)]{Popov2005}
{\sc \au{Popov, Yuri~O}} \yr{2005}  \at{Evaporative deposition patterns:
  spatial dimensions of the deposit}.  \jt{Physical Review E}  \bvol{71}~(3),
  \pg{036313}.

\bibitem[Qu{\'e}r{\'e}(2013)]{2013_Quere}
{\sc \au{Qu{\'e}r{\'e}, David}} \yr{2013}  \at{Leidenfrost dynamics}.
  \jt{Annual Review of Fluid Mechanics}  \bvol{45},  \pg{197--215}.

\bibitem[Snoeijer {\em et~al.\/}(2009)Snoeijer, Brunet \&
  Eggers]{2009_Snoeijer}
{\sc \au{Snoeijer, J.~H.}, \au{Brunet, P.} \& \au{Eggers, J.}} \yr{2009}
  \at{Maximum size of drops levitated by an air cushion}.  \jt{Phys. Rev. E}
  \bvol{79},  \pg{036307}.

\bibitem[Sobac {\em et~al.\/}(2014)Sobac, Rednikov, Dorbolo \&
  Colinet]{2014_Sobac}
{\sc \au{Sobac, Benjamin}, \au{Rednikov, Alexey}, \au{Dorbolo, St{\'e}phane} \&
  \au{Colinet, Pierre}} \yr{2014}  \at{Leidenfrost effect: Accurate drop shape
  modeling and refined scaling laws}.  \jt{Physical Review E}  \bvol{90}~(5),
  \pg{053011}.

\bibitem[Sobac {\em et~al.\/}(2017)Sobac, Rednikov, Dorbolo \&
  Colinet]{2017_Sobac}
{\sc \au{Sobac, Benjamin}, \au{Rednikov, Alexey}, \au{Dorbolo, St{\'e}phane} \&
  \au{Colinet, Pierre}} \yr{2017}  \at{Self-propelled leidenfrost drops on a
  thermal gradient: A theoretical study}.  \jt{Physics of Fluids}
  \bvol{29}~(8).

\bibitem[Sobac {\em et~al.\/}(2021)Sobac, Rednikov, Dorbolo \&
  Colinet]{2021_Sobac_Erratum}
{\sc \au{Sobac, Benjamin}, \au{Rednikov, Alexei}, \au{Dorbolo, St{\'e}phane} \&
  \au{Colinet, Pierre}} \yr{2021}  \at{Erratum: Leidenfrost effect: Accurate
  drop shape modeling and refined scaling laws [phys. rev. e 90, 053011
  (2014)]}.  \jt{Physical Review E}  \bvol{103}~(3),  \pg{039901}.

\bibitem[Sobac {\em et~al.\/}(2015)Sobac, Talbot, Haut, Rednikov \&
  Colinet]{2015_Sobac}
{\sc \au{Sobac, Benjamin}, \au{Talbot, Pauline}, \au{Haut, Beno{\^\i}t},
  \au{Rednikov, Alexei} \& \au{Colinet, Pierre}} \yr{2015}  \at{A comprehensive
  analysis of the evaporation of a liquid spherical drop}.  \jt{Journal of
  colloid and interface science}  \bvol{438},  \pg{306--317}.

\bibitem[Soto {\em et~al.\/}(2016)Soto, Lagubeau, Clanet \&
  Qu{\'e}r{\'e}]{2016_Soto}
{\sc \au{Soto, Dan}, \au{Lagubeau, Guillaume}, \au{Clanet, Christophe} \&
  \au{Qu{\'e}r{\'e}, David}} \yr{2016}  \at{Surfing on a herringbone}.
  \jt{Physical Review Fluids}  \bvol{1}~(1),  \pg{013902}.

\bibitem[Stewart(2022)]{2022_Stewart}
{\sc \au{Stewart, Se{\'a}n~M}} \yr{2022}  \at{Leidenfrost drop dynamics: a
  forgotten past and modern day rediscoveries}.  \jt{European Journal of
  Physics}  \bvol{43}~(2),  \pg{023001}.

\end{thebibliography}

\end{document}